\documentclass[aps,prd,twocolumn,superscriptaddress,showkeys,floatfix,amsmath,amssymb,amsfonts,longbibliography]{revtex4-2}
\usepackage[dvipsnames]{xcolor}
\usepackage{amsmath}
\usepackage{dcolumn}
\usepackage{lipsum}
\usepackage{amssymb}
\usepackage{soul}
\usepackage{url}
\usepackage{epsfig}
\usepackage{graphicx}
\usepackage{amsmath}
\usepackage{bm}
\usepackage{setspace}
\usepackage{appendix}
\usepackage{amsthm}
\usepackage{bbold}
\usepackage{dcolumn}
\usepackage{epsfig}
\usepackage{graphics}
\usepackage{graphicx}
\usepackage[utf8]{inputenc}
\usepackage{tikz}
\usepackage{tikz-feynman}
\tikzfeynmanset{compat=1.1.0}

\usepackage{natbib}
\usepackage{graphicx}
\usepackage{dcolumn}
\usepackage{bm}
\usepackage{amsmath}
\usepackage{float}
\usepackage{multirow}
\usepackage{slashed}
\usepackage{booktabs}
\usepackage{array}
\usepackage{tabularx}
\usepackage{xcolor}
\usepackage{physics}
\usepackage{multirow}
\usepackage{gensymb}
\usepackage{mathtools,braket}
\usepackage{subcaption}
\usepackage{lipsum}  
\usepackage{color}
\usepackage{soul}
\usepackage{placeins}
\usepackage[colorlinks=true]{hyperref}
\usepackage{makecell}
\usepackage{booktabs}
\usepackage{multirow}
\usepackage{tabularx}
\usepackage{makecell}
\renewcommand{\arraystretch}{1.25}
\usepackage{bm}
\usepackage[normalem]{ulem}
\usepackage{xspace}
\usepackage{cancel}
\usepackage{float}
\usepackage{multirow}
\usepackage{lineno}
\definecolor{darkgreen}{rgb}{0,0.5,0}
\definecolor{purple}{rgb}{0.5,0,0.5}
\definecolor{nblue}{rgb}{0.0,0.0,0.50}
\definecolor{scarlet}{rgb}{1.0,0.2,0}
\definecolor{darkmagenta}{rgb}{0.55, 0.0, 0.55}
\definecolor{darkolivegreen}{rgb}{0.33, 0.42, 0.18}
\definecolor{darkcandyapplered}{rgb}{0.64, 0.0, 0.0}

\hypersetup{
    colorlinks=true,
    linkcolor=purple,
    citecolor=purple,
    urlcolor=blue
}
\newcommand{\be}{\begin{equation}}

\newcommand{\ee}{\end{equation}}
\newcommand{\bea}{\begin{eqnarray}}
\newcommand{\eea}{\end{eqnarray}}
\newcommand{\beas}{\begin{eqnarray*}}
\newcommand{\eeas}{\end{eqnarray*}}

\usepackage{color}

\usepackage[normalem]{ulem} 
\renewcommand\sout{\bgroup \color[rgb]{0.55,0.00,0.99} \ULdepth=-.5ex \ULset}

\usepackage{orcidlink}
\usepackage{listings}

\begin{document}

\title{Leading-Neutron Electroproduction at HERA and the EIC: Sullivan
  Process, Target Fragmentation, and Pion PDFs}

\author{Wen-Chen Chang, \orcidlink{0000-0002-1695-7830}}
\email{changwc@phys.sinica.edu.tw}
\affiliation{Institute of Physics, Academia Sinica, Taipei 11529, Taiwan}

\author{Chia-Yu Hsieh, \orcidlink{0009-0002-3968-1985}}
\email{cyhsieh@phys.sinica.edu.tw}
\affiliation{Institute of Physics, Academia Sinica, Taipei 11529, Taiwan}

\author{Satyajit Puhan, \orcidlink{0009-0004-9766-5005}}
\email{puhansatyajit@gmail.com}
\affiliation{Institute of Physics, Academia Sinica, Taipei 11529, Taiwan}

\begin{abstract}

Leading-neutron electroproduction measurements from the H1 and ZEUS
experiments at HERA have been used to constrain pion parton
distribution functions (PDFs) at small momentum fractions within the
Sullivan one-pion-exchange (OPE) framework, complementing
large-$x_\pi$ constraints from pion-induced Drell--Yan
measurements. Previous analyses have focused primarily on the region
of large neutron longitudinal momentum fraction, $x_L$, where
contributions from deep-inelastic scattering (DIS) target
fragmentation are suppressed. In this work, we use the \textsc{Pythia}
event generator to model the target-fragmentation contribution and
show that its combination with the OPE contribution reproduces the
main features of the HERA leading-neutron data over the full measured
$x_L$ range without introducing additional ad hoc normalization
factors. This result demonstrates the potential of incorporating a
broader range of leading-neutron data into future global analyses,
thereby extending sensitivity to smaller pion momentum fractions
$x_\pi$. We investigate the model dependence associated with the
pion--nucleon vertex form factor and show that the HERA data are
sensitive to different target-fragmentation treatments implemented in
\textsc{Pythia}. Finally, we present projections for leading-neutron
production at the future U.S. Electron-Ion Collider (EIC), identifying
beam-energy configurations and kinematic regions that provide enhanced
sensitivity to pion structure while suppressing DIS
target-fragmentation contributions.

\end{abstract}

\maketitle

\section{Introduction}
\label{sec:intro}

The pion, as the Nambu--Goldstone boson associated with the dynamical
chiral symmetry breaking of the strong interaction, is the lightest
QCD bound state.  Because of its small mass, the pion plays a dominant
role in the long-range nucleon--nucleon interaction. Understanding the
internal structure of the pion is therefore crucial for investigating
the non-perturbative regime of QCD~\cite{Horn:2016rip}. Since
experiments involving scattering off a stationary pion target are not
feasible, current knowledge of pion PDFs relies primarily on
pion-induced Drell--Yan data~\cite{Chang:2013opa}. While Drell--Yan
processes enable the determination of pion valence-quark distributions
at $x > 0.2$, additional experimental inputs are required to constrain
the sea-quark and gluon distributions, particularly in the small-$x$
regime. Proposed complementary processes include prompt-photon
production~\cite{Novikov:2020snp}, charmonium
production~\cite{Gluck:1977zm, Barger:1980mg}, and DIS leading-neutron
production~\cite{Barry:2018ort}.

Prompt-photon production, $\pi N \rightarrow \gamma
X$~\cite{WA70:1987bai,Gordon:1993qc}, provides a means to constrain
the gluon content of the pion through the Compton subprocess $qg
\rightarrow \gamma q$ at leading order (LO) and $qg \rightarrow \gamma
qg$ at next-to-leading order (NLO). The theoretical description of
this process is relatively robust. However, the measured cross
sections suffer from sizable experimental uncertainties due to
contamination from photons originating from $\pi^0$ decays.

Pion-induced heavy-quarkonium production, such as $J/\psi$ and
$\Upsilon$~\cite{Barger:1980mg} production, offers another potential
probe. The production cross sections are relatively large, and the
dimuon decay channels are experimentally clean, resulting in a
substantial amount of available
data~\cite{Gavai:1994in,Schuler:1996ku}. While the theoretical
calculation of heavy-quark pair $Q\bar{Q}$ production at the partonic
level is well established, the model dependence arises in the
transition of the $Q\bar{Q}$ pair into a physical quarkonium
state. Recently, this model dependence in pion-induced charmonium
production has been investigated using the color evaporation model
(CEM) and non-relativistic QCD (NRQCD)~\cite{Chang:2020rdy,
  Hsieh:2021yzg, Chang:2022pcb}. These studies demonstrate that,
despite model-dependent uncertainties, charmonium production retains
good sensitivity to the gluon content of the pion.

Due to the forward acceptance limitations of fixed-target experiments,
pion-induced processes such as Drell-Yan, prompt-photon, and
charmonium production constrain pion PDFs primarily at large momentum
fractions ($x > 0.2$). The primary accessible channel for constraining
small-$x$ pion PDFs is leading-neutron production in deep-inelastic
scattering (LN-DIS), measured by the H1~\cite{H1:2010hym} and
ZEUS~\cite{ZEUS:2002gig} collaborations at HERA. This process has been
studies within several theoretical frameworks, including Reggeon
exchange~\cite{Bishari:1972tx}, the color dipole
formalism~\cite{Kopeliovich:2012fd}, nucleon fracture
functions~\cite{deFlorian:1997wi}, and
OPE~\cite{Holtmann:1994rs}. Under the Sullivan process
picture~\cite{Sullivan:1971kd}---where the incoming proton fluctuates
into a virtual pion cloud---the emitted pion serves as an effective
target to extract pion structure functions via
OPE~\cite{Khoze:2006hw,McKenney:2015xis}. Indeed, H1 and ZEUS data
have been included into recent global QCD analyses by the
JAM~\cite{Barry:2018ort,Cao:2021aci,Barry:2021osv} and
FantoPDF~\cite{Kotz:2023pbu} teams, constraining sea-quark and gluon
distributions down to $x \approx 0.001$. However, this extraction is
known to be subject to notable systematic uncertainties stemming from
the off-shell nature of the virtual pion in the fluctuated Fock
state~\cite{Qin:2017lcd} and the modeling of the pion--nucleon
vertex~\cite{McKenney:2015xis}. Furthermore, non-OPE DIS target
fragmentation also contributes to leading-neutron production. To
suppress possible contamination from target fragmentation, analyses
such as the JAM pion PDF fits explicitly restricted the data to the
highest neutron longitudinal-momentum bins, typically $x_L\gtrsim0.8$.

In the H1 analysis~\cite{H1:2010hym}, the leading-neutron data were
described by combining inclusive DIS events generated with
\textsc{DJANGO}, which accounts for proton-remnant fragmentation, with
the OPE contribution simulated using \textsc{RAPGAP}. A good
description of the data required ad hoc normalization factors of 1.2
and 0.65 for the target-fragmentation (\textsc{DJANGO}) and OPE
(\textsc{RAPGAP}-$\pi$) contributions, respectively. In this work, we
revisit this issue and investigate whether the H1 and ZEUS LN-DIS data
over the full measured neutron-momentum range can be described by the
combined contributions from generic DIS target fragmentation and the
Sullivan process without such rescaling. The target-fragmentation
contribution is modeled using the \textsc{Pythia} event generator with
Lund string fragmentation, while the Sullivan contribution is
calculated within chiral effective theory~\cite{McKenney:2015xis}
using three pion PDF sets: JAM, xFitter, and GRV. We systematically
assess the model dependence associated with both the \textsc{Pythia}
target-fragmentation treatment and the pion--nucleon vertex form
factor. Our results indicate that the combined framework nicely
reproduces the main features of the HERA data over a broad kinematic
range, suggesting that a larger fraction of the available
leading-neutron measurements could be incorporated into future global
analyses to improve constraints on pion PDFs.

The future U.S. Electron-Ion Collider
(EIC)~\cite{Accardi:2012qut,AbdulKhalek:2021gbh} will substantially
extend the kinematic reach and statistical precision of DIS
leading-neutron measurements, with luminosities approximately three
orders of magnitude higher than those achieved at HERA. Its integrated
far-forward detector system, including a Zero-Degree Calorimeter
(ZDC), Roman Pots, and off-momentum detectors, will provide extensive
acceptance for forward-going particles. The combination of high
luminosity and broad forward acceptance will enable multidimensional
measurements of the leading-neutron production with significantly
improved statistical precision, providing sensitivity to pion PDFs
over a substantially extended kinematic range. These measurements will
complement fixed-target programs at Jefferson Lab~\cite{TDIS:2014} and
COMPASS++/AMBER~\cite{Bernhard:2019jqz}, as well as future collider
measurements at EicC~\cite{Xie:2020uck,Lu:2025bnm}.

The paper is organized as follows. Section~\ref{sec:DIS-LN} introduces
the kinematic variables used in measurements of DIS leading-neutron
production. In Sec.~\ref{sec:Sullivan}, we present the formalism for
the OPE Sullivan process, including the pion flux and pion structure
functions. Section~\ref{sec:Data_Sullivan} compares the H1 and ZEUS
leading-neutron data with calculations based on the OPE Sullivan
process using various pion-flux parameterizations and pion PDF
sets. In Sec.~\ref{sec:DIS_fragmentation}, we describe the DIS
target-fragmentation models implemented in
\textsc{Pythia}~8. Section~\ref{sec:Data_CombinedResults} compares the
experimental measurements with the unscaled combined contributions
from the Sullivan process and target
fragmentation. Section~\ref{sec:EIC} presents predictions for the
differential cross sections and kinematic coverage of leading-neutron
production for three beam-energy configurations at the future
U.S. EIC. Finally, Sec.~\ref{sec:summary} summarizes our main findings
and presents a brief outlook.

\section{DIS Leading-Neutron Production}
\label{sec:DIS-LN}

\begin{figure*}[!htbp]
\centering
\includegraphics[width=0.8\linewidth]{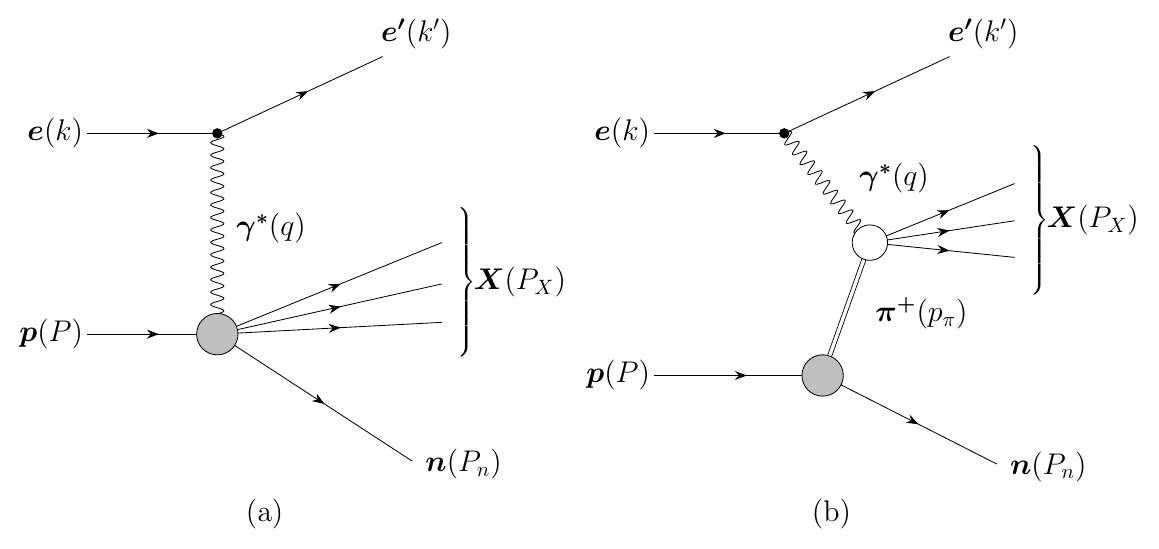}
\caption{(a) Schematic diagram of leading-neutron production through
  proton target fragmentation in DIS. (b) Schematic diagram of the
  Sullivan process with one-pion exchange. The four-momenta are
  denoted by $k$ ($k'$) for the incoming (scattered) electron, $P$ for
  the incoming proton, $q=k-k'$ for the exchanged virtual photon,
  $P_n$ for the outgoing neutron, $P_X$ for the hadronic final state,
  and $p_\pi=P-P_n$ for the exchanged pion.}
\label{fig:sullivan}
\end{figure*}

Semi-inclusive LN-DIS, $e(k) + p(P) \to e(k') + n(P_n) + X(P_X)$, is
measured by tagging a forward-moving neutron emitted along the
direction of the incoming proton beam. Figure~\ref{fig:sullivan}(a)
illustrates the generic mechanism in which the neutron originates from
proton target fragmentation. An incoming electron with four-momentum
$k$ scatters from a proton with four-momentum $P$ through the exchange
of a virtual photon with momentum $q=k-k'$, where $k'$ denotes the
four-momentum of the scattered electron. The final-state neutron, with
four-momentum $P_n$, is detected at forward angles, while $X(P_X)$
denotes the remaining undetected hadronic system.

The nominal inclusive DIS kinematics are characterized by three
variables, namely the photon virtuality $Q^2$, the Bjorken scaling
variable $x_B$, and the inelasticity $y$, which are defined as
\begin{equation}
Q^2 \equiv -q^2, \quad
x_B = \frac{Q^2}{2P \cdot q}, \quad
y = \frac{P \cdot q}{P \cdot k}.
\end{equation}
These variables are related to the squared center-of-mass energy,
$s=(P+k)^2$, through
\begin{equation}
Q^2 \approx s x_B y,
\end{equation}
where the approximation corresponds to neglecting the initial-state
particle masses. The invariant mass squared of the hadronic final
state is given by
\begin{equation}
W^2 = (P+q)^2 = m_p^2 + \frac{Q^2(1-x_B)}{x_B},
\end{equation}
where $m_p$ denotes the proton mass.

With the detection of the forward neutron in the final state,
semi-inclusive leading-neutron production is characterized by two
additional kinematic variables: the neutron longitudinal momentum
fraction $x_L$ and the squared four-momentum transfer $t$. The
longitudinal momentum fraction is defined as
\begin{equation}
x_L = \frac{P_n \cdot k}{P \cdot k} \approx \frac{E_n}{E_p},
\end{equation}
and represents the fraction of the initial proton beam energy carried
by the outgoing neutron. The squared four-momentum transfer between
the initial proton and the final-state neutron is given by
\begin{equation}
t = (P-P_n)^2 \approx -\frac{p_T^2}{x_L} -
(1-x_L)\left(\frac{m_n^2}{x_L}-m_p^2\right),
\label{eq:t}
\end{equation}
where $E_p$ ($m_p$) and $E_n$ ($m_n$) denote the energy (mass) of the
initial proton and outgoing neutron, respectively, and $p_T$ is the
transverse momentum of the neutron. Eq.~(\ref{eq:t}) shows that $t$
depends on both $x_L$ and $p_T$. Thus, $(x_L,t)$ and $(x_L,p_T)$
provide equivalent descriptions of the neutron kinematics. For a given
$x_L$, the maximum value of $t$ corresponds to $p_T=0$ and is given by
$t_{\rm max}=-(1-x_L)(m_n^2/x_L-m_p^2)$.

The four-fold differential cross section for leading-neutron
production can be parameterized in terms of the semi-inclusive
structure function $F_{2}^{LN(4)}$, defined as~\cite{H1:2010hym}
\begin{equation}
\begin{aligned}
\frac{d^{4}\sigma(ep \rightarrow e'nX)}{dx_B \, dQ^{2} \, dx_{L} \, dt}
&= \frac{4\pi\alpha^{2}}{x_B Q^{4}} \left(1-y+\frac{y^{2}}{2}\right) \\
&\quad \times F_{2}^{\text{LN}(4)}(x_B, Q^2, x_{L}, t) .
\end{aligned}
\label{eq:LN_D4}
\end{equation}
where $x_B$, $Q^2$, and $y$ are the usual DIS variables, while
$x_L$ and $t$ characterize the kinematics of the leading neutron.

Due to detector acceptance, neutron measurements are restricted to
polar angles $\theta\leq\theta_{\max}$, corresponding to a maximum
accessible transverse momentum $p_{T,\max}\approx E_p
x_L\theta_{\max}$. Integrating the four-fold differential cross
section over the accepted $p_T$ range, or equivalently over $t$,
yields the three-fold differential cross section,
\begin{equation}
\begin{aligned}
\frac{d^3\sigma(ep\to e'nX)} {dx_B dQ^2 dx_L} &= \int_{t_0}^{t_{\max}}
\frac{d^4\sigma(ep\to e'nX)} {dx_B dQ^2 dx_L dt} dt \\
&= \frac{4\pi\alpha^2}{x_BQ^4} \left(1-y+\frac{y^2}{2}\right) \\ 
&\quad\times F_2^{\mathrm{LN}(3)}(x_B,Q^2,x_L),
\end{aligned}
\label{eq:ln_cross_section}
\end{equation}
where $t_{\max}\equiv t(p_T=0)$ is the maximum (least negative) value
of $t$, corresponding to the minimum momentum-transfer magnitude,
while $t_0\equiv t(p_{T}^{\max})$ is determined by the maximum accepted
neutron transverse momentum. The quantity $F_2^{\mathrm{LN}(3)}$
denotes the three-fold leading-neutron structure function obtained by
integrating $F_2^{\mathrm{LN}(4)}$ over the accepted $t$ range.

Previous studies have shown that leading-neutron production at large
longitudinal momentum fraction ($x_L \gtrsim 0.6$) and small
transverse momentum can be successfully described by the OPE
mechanism~\cite{H1:2010hym,ZEUS:2002gig}, as illustrated by
Fig.~\ref{fig:sullivan}(b) and discussed further in
Sec.~\ref{sec:Sullivan}. In this framework, the exchanged virtual pion
acts as an effective target probed by the virtual photon. The
corresponding Bjorken variable for the pion is
\begin{equation}
x_{\pi} = \frac{Q^{2}}{2p_{\pi}\cdot q} \simeq \frac{x_{B}}{1-x_{L}},
\label{eq:xpixB}
\end{equation}
where $p_\pi=P-P_n$ is the four-momentum of the exchanged pion. The
variable $x_\pi$ therefore represents the momentum fraction of the
struck parton in the pion, analogous to the role of $x_B$ in
inclusive DIS on the proton. Consequently, measurements of the
semi-inclusive process $ep\to e'nX$ provide experimental access to
the partonic structure of the pion through the leading-neutron
structure function $F_{2}^{LN(3)}(x_B,Q^{2},x_L)$.

\section{Modeling of one-pion exchange (OPE)}
\label{sec:Sullivan}

The theoretical description of leading-neutron production within the
OPE framework relies on the factorization of the nonperturbative
proton-to-neutron pion-emission process from the hard lepton--pion
scattering subprocess. In chiral effective field
theory~\cite{Burkardt:2012hk,McKenney:2015xis}, the semi-inclusive
leading-neutron structure function can be expressed as the product of
the proton-to-neutron splitting function $f_{\pi N}$ and the pion
structure function $F_2^\pi$ as
\begin{equation}
F_2^{\mathrm{LN}(4)}(x_B,Q^2,x_L,t) = 2 f_{\pi N}(x_L,k_\perp) F_2^\pi(x_\pi,Q^2),
\label{eq:f2ln}
\end{equation}
where, $k_\perp$ denotes the transverse momentum of the exchanged
pion. In the OPE picture, transverse-momentum conservation gives
$\boldsymbol{k}_\perp=-\boldsymbol{p}_T$, so that $k_\perp=p_T$ in
magnitude. The pion structure function $F_2^\pi$ is assume to be
independent of $k_\perp$. In the approximation $m_n\simeq m_p$, the
pion virtuality $t$ can be reconstructed from the measured energy and
transverse momentum of the leading neutron via $t = -[k_\perp^2 +
  (1-x_L)^2 m_p^2]/x_L$ (Eq.~(\ref{eq:t})). The factor of $2$ in
Eq.~(\ref{eq:f2ln}) arises from the isospin relation associated with
the $p\to n\pi^+$ fluctuation.

Because $F_2^\pi$ is assumed to be independent of $t$ (or $p_T$),
integrating $F_2^{\text{LN}(4)}$ over $t$ is equivalent to integrating
the pion flux factor $f_{\pi N}(x_L, k_\perp)$ over $k_\perp$.
Consequently, the $p_T$-integrated structure function
$F_2^{\text{LN}(3)}(x_B, Q^2, x_L)$ in Eq.~(\ref{eq:ln_cross_section})
can be written as
\begin{equation}
F_2^{\text{LN}(3)}(x_B, Q^2, x_L) = 2 f_{\pi N}(x_L) F_2^{\pi}(x_{\pi}, Q^2),
\label{eq:F2N_OPE}
\end{equation}
where the integrated pion flux $f_{\pi N}(x_L)$ is defined by
\begin{equation}
f_{\pi N}(x_L) = \int dk_\perp^2 \, f_{\pi N}(x_L, k_\perp).
\end{equation}

\subsection{Pion Flux}
\label{pionflux}

In chiral effective field theory, the pion flux $f_{\pi N}(x_L)$ can
be written as~\cite{Burkardt:2012hk,Salamu:2014pka,McKenney:2015xis}
\begin{equation}
\begin{aligned}
f_{\pi N}(x_L) &= \frac{g_A^2 M_N^2}{(4\pi f_\pi)^2} \int_0^{\infty}
d k_\perp^2 \, \frac{1-x_L}{x_L^2 D_{\pi N}^2} \\
&\quad \times \left[ k_\perp^2 + (1-x_L)^2 M_N^2 \right] 
|F(k_\perp^2, x_L)|^2,
\end{aligned}
\label{eq:flux_general}
\end{equation}
where $M_N$ is the nucleon mass, $g_A$ is the nucleon axial-vector
coupling, and $f_\pi$ is the pion decay constant. The energy
denominator is given by
\begin{equation}
D_{\pi N} = -\frac{k_\perp^2+(1-x_L)^2M_N^2+x_Lm_\pi^2}{x_L},
\end{equation}
and describes the virtuality of the intermediate $\pi N$ state. The
function $F(k_\perp^2,x_L)$ is a phenomenological regulator that
suppresses contributions from large $k_\perp$ and highly off-shell
configurations and parametrizes the short-distance behavior of the
$\pi N$ interaction.

Following the theoretical frameworks of
Refs.~\cite{McKenney:2015xis,Barry:2018ort}, we consider five form
factor parameterizations for the regulator $F(k_\perp^2, x_L)$, which
are expressed as
\begin{equation}
F = 
\begin{cases}
\exp\!\left(\dfrac{M_N^2 - s}{\Lambda^2}\right), & \text{(i) } s\text{-exp.} \\[1.8ex]
\exp\!\left(\dfrac{D_{\pi N}}{\Lambda^2}\right), & \text{(ii) } t\text{-exp.} \\[1.8ex]
\dfrac{\Lambda^2 - m_\pi^2}{\Lambda^2 - t}, & \text{(iii) Monopole} \\[1.8ex]
(1 - x_L)^{-\alpha_\pi(t)} \exp\!\left(\dfrac{D_{\pi N}}{\Lambda^2}\right), & \text{(iv) Regge} \\[1.8ex]
\left[ 1 - \dfrac{D_{\pi N}^2}{(\Lambda^2 - t)^2} \right]^{1/2}, & \text{(v) PV}
\end{cases}
\label{eq:regulators}
\end{equation}
where $s=(k_\perp^2+M_N^2)/x_L+(k_\perp^2+m_\pi^2)/(1-x_L)$ is the
invariant mass squared of the $\pi N$ system,
$\alpha_\pi(t)=\alpha_\pi' t$ is the pion Regge trajectory with slope
parameter $\alpha_\pi'$, and $\Lambda$ denotes the cutoff scale. The
physical constants and cutoff parameters adopted from the JAM18
analysis~\cite{Barry:2018ort} are listed in
Table~\ref{tab:flux_parameters}.

\begin{table}[!htbp]
\centering
\caption{Physical constants and pion-flux model parameters adopted
  from the JAM18 analysis~\cite{Barry:2018ort}.}
\label{tab:flux_parameters}
\begin{tabular}{lll}
\hline\hline
Parameter & Value & Description \\
\hline
$g_A$ & $1.267$ & Axial-vector coupling constant \\
$f_\pi$ & $0.093~\text{GeV}$ & Pion decay constant \\
$\alpha_\pi'$ & $1.0~\text{GeV}^{-2}$ & Regge trajectory slope \\
\hline
$\Lambda_{s\text{-exp}}$ & $1.31~\text{GeV}$ & $s$-dependent exponential cutoff \\
$\Lambda_{t\text{-exp}}$ & $0.58~\text{GeV}$ & $t$-dependent exponential cutoff \\
$\Lambda_{\text{mono}}$ & $0.52~\text{GeV}$ & Monopole cutoff \\
$\Lambda_{\text{Regge}}$ & $0.78~\text{GeV}$ & Regge cutoff \\
$\Lambda_{\text{PV}}$ & $0.25~\text{GeV}$ & Pauli--Villars cutoff \\
\hline\hline
\end{tabular}
\end{table}

\begin{figure*}[t]
\centering
\includegraphics[width=\textwidth]{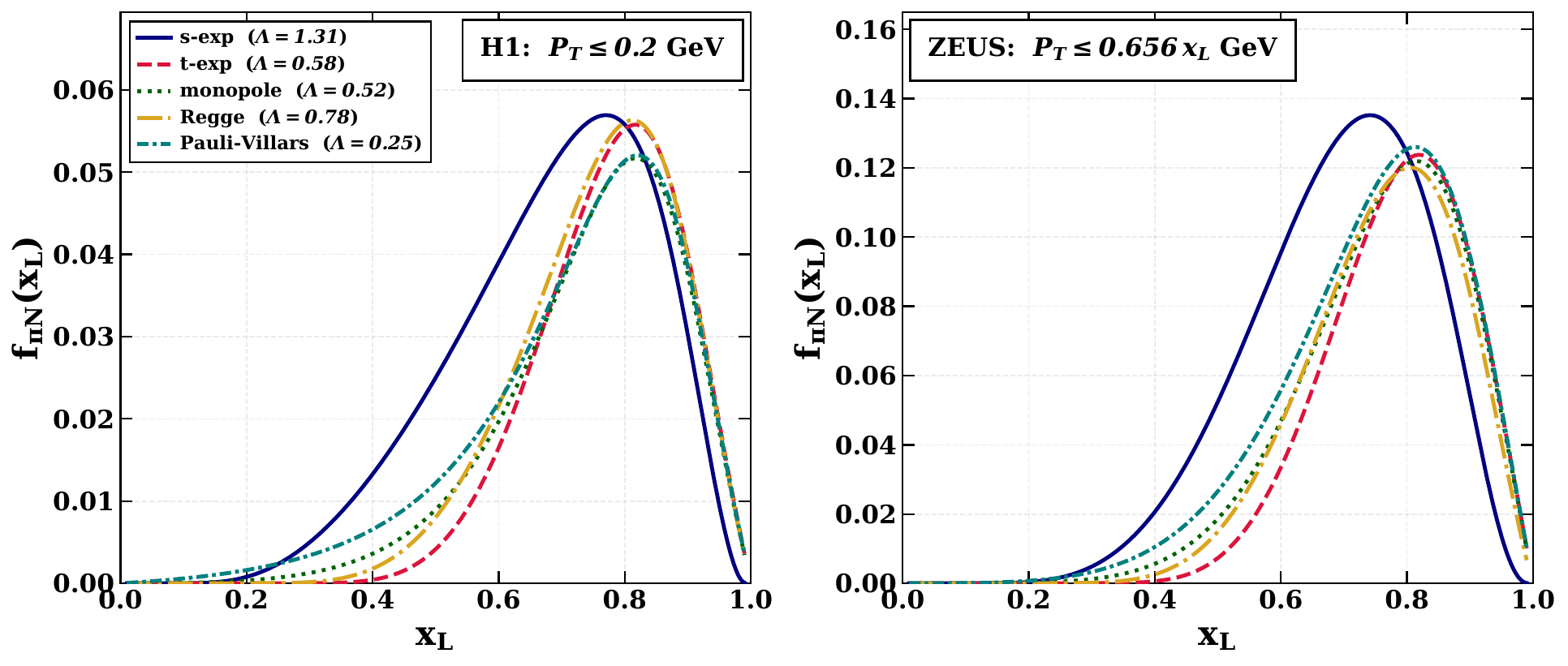}
\caption{Comparison of the pion fluxes $f_{\pi N}(x_L)$ obtained using
  the $s$-exponential, $t$-exponential, $t$-monopole, Reggeized, and
  Pauli--Villars form factor parameterizations under H1 and ZEUS
  kinematics.  The left and right panels correspond to the
  transverse-momentum acceptance cuts $p_T \le 0.2~\mathrm{GeV}$
  (H1) and $p_T \le 0.656\,x_L~\mathrm{GeV}$ (ZEUS),
  respectively. The cutoff parameters $\Lambda$ for each pion flux
  model are adopted from the JAM18 global fit~\cite{Barry:2018ort}.}
\label{fig:pionflux}
\end{figure*}

Figure~\ref{fig:pionflux} shows the pion fluxes $f_{\pi N}(x_L)$ for
these five regulator models evaluated within the fiducial
transverse-momentum acceptances of H1 ($p_T \leq 0.20~\mathrm{GeV}$)
and ZEUS ($p_T \leq 0.656 x_L~\mathrm{GeV}$). Despite differences in
normalization and detailed $x_L$ dependence, all models exhibit a
similar overall behavior: the flux increases from $x_L \sim 0.3$--0.5,
reaches a broad maximum around $x_L \sim 0.6$--0.85, and decreases
rapidly as $x_L \to 1$. This behavior reflects the interplay between
the pion propagator, the proton-to-neutron splitting kinematics, and
the regulator dependence of the $\pi N$ vertex. At smaller $x_L$, the
exchanged pion generally carries larger spacelike virtuality,
suppressing the pion-exchange contribution. With increasing $x_L$, the
magnitude of the minimum pion virtuality decreases, enhancing the
contribution from the pion-pole region. As $x_L\to1$, however, the
longitudinal momentum carried by the exchanged pion vanishes, leading
to a strong kinematic suppression of the pion flux.

Among the models considered, the $s$-dependent exponential regulator
produces a noticeably softer and broader $x_L$ distribution than the
other parameterizations. In particular, for $x_L \gtrsim 0.6$, the
corresponding pion flux $f_{\pi N}(x_L)$ is systematically smaller
than those predicted by the other four models. This feature is
essential in understanding the results of the comparison of
large-$x_L$ data and OPE Sullivan contribution in
Sec.~\ref{sec:Data_Sullivan}.

Beyond the simple OPE picture, absorptive rescattering
corrections~\cite{Nikolaev:1997cnn, Nikolaev:1998jj, DAlesio:1998uav,
  Khoze:2006hw} and non-pionic contributions~\cite{Khoze:2006hw,
  Kopeliovich:2012fd} have also been investigated. However, these
effects are not included in the present analysis.

\subsection{Pion Structure functions}
\label{pionSF}

Given a set of pion PDFs, the corresponding pion structure function
$F_2^\pi(x_\pi, Q^2)$ at NLO accuracy in the $\overline{\text{MS}}$
scheme is constructed by convoluting the pion quark and gluon PDFs
with the corresponding Wilson coefficient functions~\cite{Ellis:1996}
as
\begin{equation}
\begin{aligned}
F_{2}^{\pi}(x_\pi, Q^{2}) &= x_\pi \sum_{q} e_q^{2} \Bigg[
  q^{\pi}(x_\pi, Q^{2}) + \bar{q}^{\pi}(x_\pi, Q^{2}) \\ 
&\quad + \frac{\alpha_s(Q^{2})}{2\pi} \left(
  C_q^{\overline{\text{MS}}} \otimes (q^{\pi} + \bar{q}^{\pi}) +
  C_g^{\overline{\text{MS}}} \otimes g^{\pi} \right) \!\Bigg],
\end{aligned}
\label{eq:F2pi}
\end{equation}
where the convolution integral is defined as
\begin{equation}
(C \otimes f)(x_\pi) = \int_{x_\pi}^1 \frac{d\xi}{\xi} \,
  C\!\left(\frac{x_\pi}{\xi}\right) f(\xi, Q^2).
\label{eq:conv}
\end{equation}
The NLO Wilson coefficient functions in the $\overline{\text{MS}}$
scheme are given by
\begin{align}
C_q^{\overline{\text{MS}}}(z) &= C_F \Bigg[ 2\left(
  \frac{\ln(1-z)}{1-z} \right)_+ - \frac{3}{2} \left( \frac{1}{1-z}
  \right)_+ \nonumber \\ 
&\quad - (1+z)\ln(1-z) - \frac{1+z^2}{1-z}\ln z + 3 + 2z \nonumber \\
&\quad - \left( \frac{\pi^2}{3} + \frac{9}{2} \right) \delta(1-z)
  \Bigg],
\label{eq:Cq}
\end{align}
and
\begin{equation}
C_g^{\overline{\text{MS}}}(z) = T_R \left[ \left((1-z)^2 + z^2\right)
  \ln\left(\frac{1-z}{z}\right) - 8z^2 + 8z - 1 \right],
\label{eq:Cg}
\end{equation}
where $C_F = 4/3$ and $T_R = 1/2$ are the standard $\text{SU}(3)_c$
color factors. The plus distribution $(\dots)_+$ regulates soft-gluon
singularities as $z \to 1$, while the $\delta(1-z)$ term accounts for
virtual and soft-real corrections that preserve the normalization of
the quark coefficient function.

\begin{figure}[!htbp]
\centering
\includegraphics[width=\linewidth]{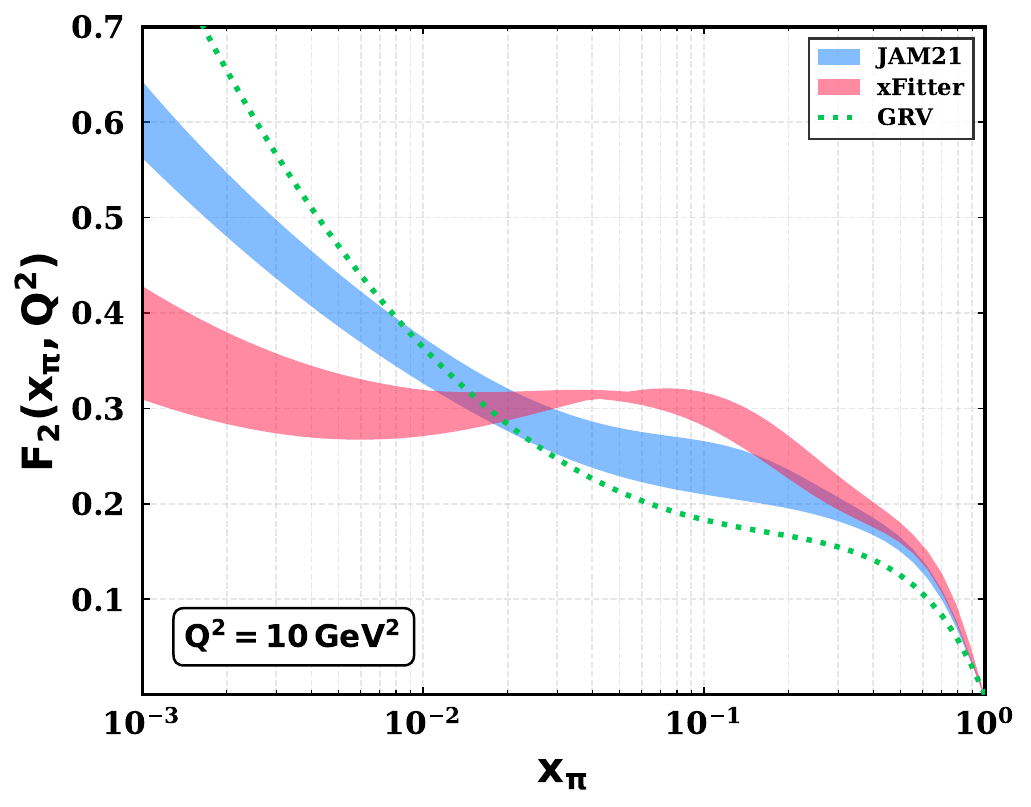}
\caption{Comparison of the pion structure function
  $F_2^\pi(x_\pi,Q^2)$ at $Q^2=10~\mathrm{GeV}^2$ obtained at NLO
  accuracy using the JAM21~\cite{Barry:2021osv},
  xFitter~\cite{Novikov:2020snp}, and GRV~\cite{Gluck:1991ey} pion
  PDFs.}
\label{fig:f2sf}
\end{figure}

In this work, we employ three sets of pion PDFs to evaluate the
leading-neutron structure functions: JAM21~\cite{Barry:2021osv},
xFitter~\cite{Novikov:2020snp}, and GRV~\cite{Gluck:1991ey}. Note that
the HERA LN-DIS measurements from H1 and ZEUS have been included in
the global analysis of the JAM21 pion PDFs along with Drell-Yan
data. Using Eqs.~(\ref{eq:F2pi})--(\ref{eq:Cg}) together with these
PDF sets, we calculate NLO $F_2^\pi(x_\pi,Q^2)$ at
$Q^2=10~\mathrm{GeV}^2$, as shown in Fig.~\ref{fig:f2sf}.

For $x_\pi \gtrsim 0.02$, the three predictions are in reasonable
agreement, reflecting the constraints provided predominantly by
pion-induced Drell-Yan data in the intermediate- and large-$x_\pi$
regions. The JAM21 and xFitter results exhibit a mild enhancement
around $x_\pi\sim0.1$--$0.2$, which is less pronounced for GRV. The
JAM21 and xFitter uncertainty bands also largely overlap in this
region, indicating that the two determinations are mutually consistent
despite differences in their fitted data sets and parametrization
assumptions. At small $x_\pi$, $x_\pi\lesssim10^{-2}$, the GRV
prediction rises substantially more rapidly than the JAM21 and xFitter
results. This behavior reflects the dynamical generation of the
sea-quark and gluon distributions in the GRV framework through QCD
evolution from a low input scale, whereas JAM21 and xFitter exhibit a
considerably flatter small-$x_\pi$ dependence.

The spread among the predictions increases toward small $x_\pi$
reflects the fact that the pion PDFs remain poorly constrained by
existing data, particularly for the sea-quark and gluon
distributions. Additional measurements in this region are therefore
important for improving our knowledge of pion structure. As
illustrated in Sec.~\ref{sec:Data_Sullivan}, leading-neutron DIS
measurements can provide unique constraints on the pion structure
function in the low-$x_\pi$ region.

\section{Comparison of H1 and ZEUS data with Sullivan-process contribution}
\label{sec:Data_Sullivan}

The H1 Collaboration~\cite{H1:2010hym} measured the leading-neutron
structure function $F_2^{\mathrm{LN}(3)}$ over the kinematic range
$1.5\times10^{-4}\leq x_B\leq3\times10^{-2}$ and $6\leq
Q^2\leq100~\mathrm{GeV}^2$, with average inelasticities $0.02\lesssim
y\lesssim0.60$ and a neutron transverse-momentum requirement
$p_T<0.2~\mathrm{GeV}$. This restrictive $p_T$ selection
preferentially probes forward neutrons at small momentum transfer,
where pion exchange is expected to provide an important
contribution. The ZEUS Collaboration~\cite{ZEUS:2002gig} covered a
substantially broader range, $8.0\times10^{-5}\leq
x_B\leq1.0\times10^{-1}$ and $4<Q^2<10^4~\mathrm{GeV}^2$, with the
neutron acceptance restricted to $\theta_n<0.8~\mathrm{mrad}$. This
angular requirement corresponds approximately to the $x_L$-dependent
limit $p_T<0.656 x_L~\mathrm{GeV}$, which is more restrictive than the
H1 requirement at sufficiently small $x_L$ but permits larger
transverse momenta at larger $x_L$. The H1 analysis additionally
imposes requirements on $y$, the hadronic invariant mass $W$, and the
scattered-electron polar angle. The experimental configurations and
kinematic selections of both measurements are summarized in
Table~\ref{tab:kinematics}.

\begin{table}[!htbp]
\centering
\caption{Experimental configurations and kinematic cuts for the
  H1~\cite{H1:2010hym} and ZEUS~\cite{ZEUS:2002gig} leading-neutron
  measurements.}
\label{tab:kinematics}
\renewcommand{\arraystretch}{1.15}
\resizebox{\columnwidth}{!}{%
\begin{tabular}{lcc}
\hline\hline
Quantity & H1 & ZEUS \\
\hline
$E_e$ ($\text{GeV}$) & $27.5$ & $27.5$ \\
$E_p$ ($\text{GeV}$) & $920$ & $820$ \\
$Q^2$ ($\text{GeV}^2$) & $6 < Q^2 < 100$ & $4 < Q^2 < 10^4$ \\
$x_B$ & $1.5 \times 10^{-4} < x_B < 3.0 \times 10^{-2}$ & $8.0 \times 10^{-5} < x_B < 1.0 \times 10^{-1}$ \\
$y$ & $0.02 < y < 0.60$ & -- \\
$W^2$ ($\text{GeV}^2$) & $> 4$ & -- \\
$E'_e$ ($\text{GeV}$) & $> 11$ & $> 10$ \\
$\theta_e$ & $156^\circ < \theta_e < 175^\circ$ & -- \\
$\theta_n$ ($\text{mrad}$) & $< 0.75$ & $< 0.80$ \\
$p_T$ ($\text{GeV}$) & $< 0.20$ & $< 0.656 \, x_L$ \\
$x_L$ & $0.32$--$0.95$ & $0.20$--$1.00$ \\
$x_L$ bins & $7$ & $12$ \\
\hline\hline
\end{tabular}%
}
\end{table}

To reduce systematic uncertainties associated with the beam
luminosity, detector acceptance, and efficiency corrections, ZEUS
measured the ratio of the semi-inclusive leading-neutron cross section
to the inclusive DIS cross section,
\begin{equation}
r(x_B,Q^2,x_L) = \frac{ d^3\sigma^{\mathrm{LN}}/ (dx_B dQ^2 dx_L)} {
  d^2\sigma^{\mathrm{inc}}/ (dx_B dQ^2)} \Delta x_L .
\label{eq:r_def}
\end{equation}
where $\Delta x_L$ denotes the $x_L$ bin width. The leading-neutron
structure function $F_2^{\mathrm{LN}(3)}$ can then be obtained using
the well-constrained proton structure function $F_2^p(x_B,Q^2)$ as
\begin{equation}
F_2^{\mathrm{LN}(3)}(x_B,Q^2,x_L)=\frac{r(x_B,Q^2,x_L)}{\Delta x_L} F_2^p(x_B,Q^2)
\label{eq:r_SF}
\end{equation}


\begin{figure*}[!htbp]
\centering
\includegraphics[width=\textwidth]{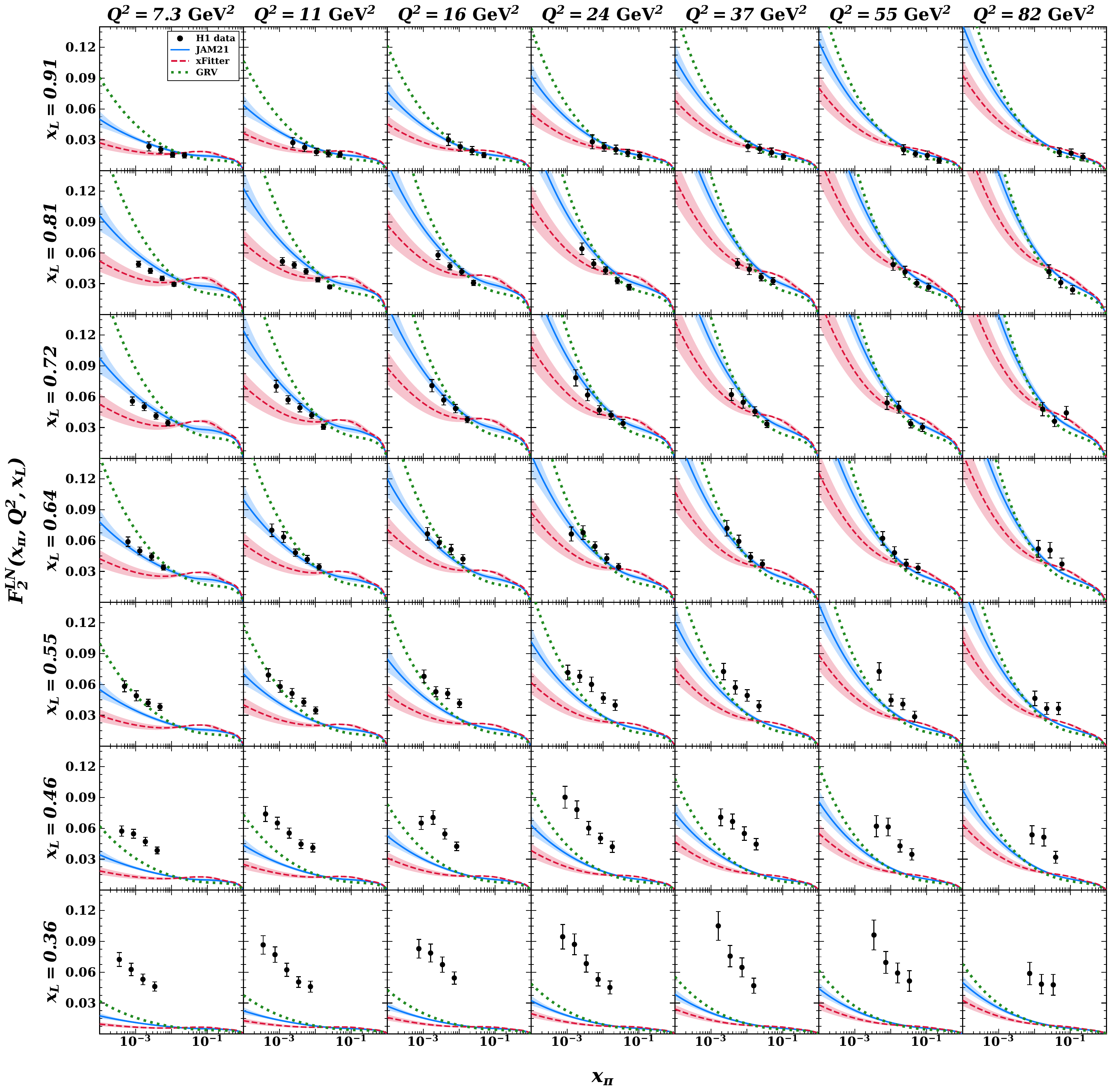}
\caption{Leading-neutron structure function
  $F_2^{\mathrm{LN}(3)}(x_\pi,Q^2,x_L)$ as a function of $x_\pi$ for
  the H1 kinematics, calculated using the $s$-dependent exponential
  pion flux. Each row corresponds to a fixed $x_L$ interval, and each
  column to a fixed $Q^2$ interval. The uncertainty bands for the
  JAM21 and xFitter predictions are obtained from the full sets of PDF
  replicas.}
\label{fig:h1f2ln}
\end{figure*}

\begin{figure*}[!hbt]
\centering
\includegraphics[width=\textwidth]{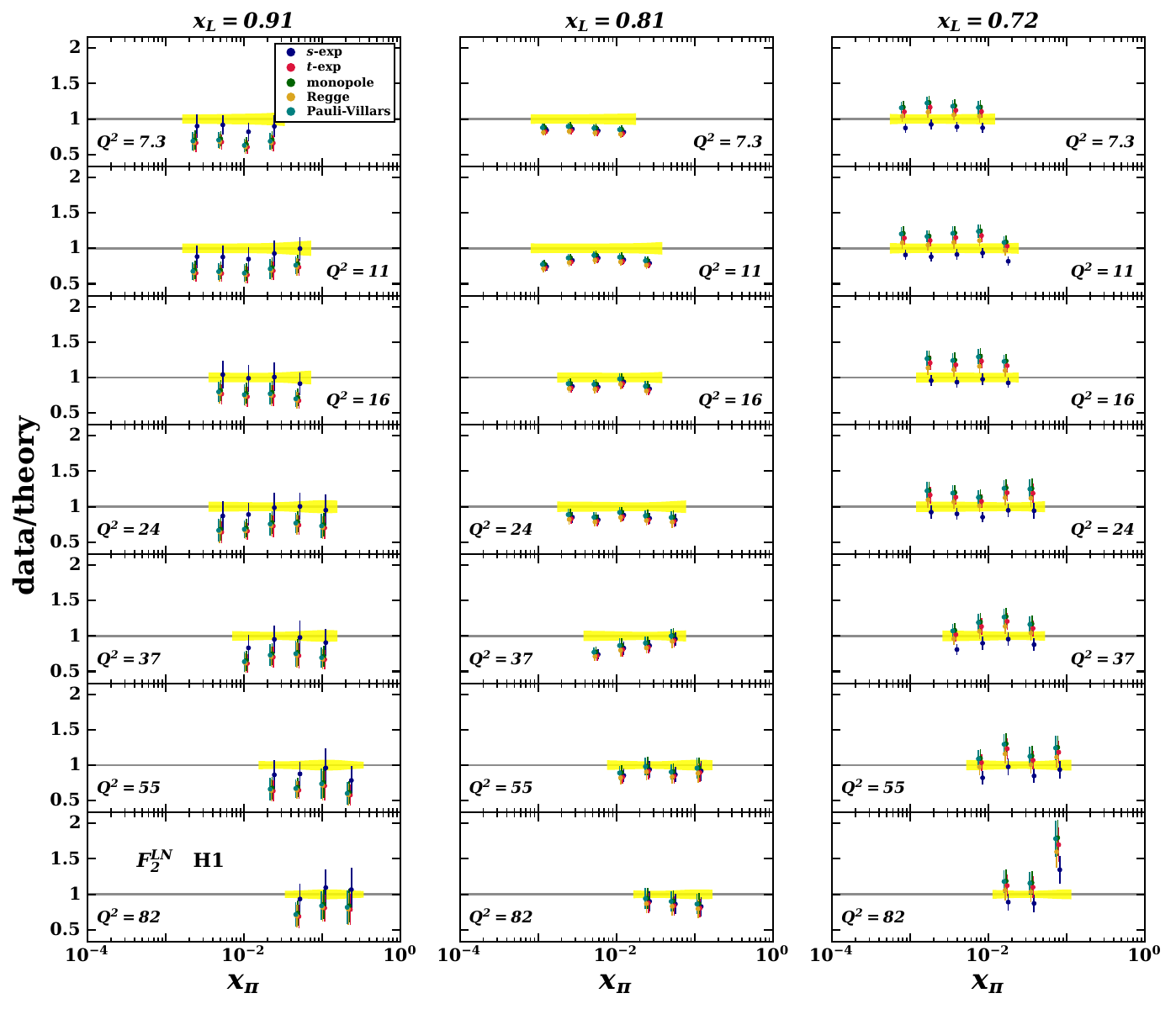}
\caption{Data-to-theory ratios for $F_2^{\mathrm{LN}(3)}$ as functions
  of $x_\pi$ in bins of $Q^2$ for the three largest H1 $x_L$ values,
  $x_L=0.91$, 0.81, and 0.72~\cite{H1:2010hym}. The theoretical
  calculations use the JAM21 pion PDF replicas together with the five
  pion-flux regulator parameterizations defined in
  Eq.~(\ref{eq:regulators}). The shaded bands represent the
  corresponding $1\sigma$ PDF uncertainties.}
\label{fig:h1datatotheoryfive}
\end{figure*}

Using the pion-flux models and pion structure functions introduced in
Sec.~\ref{sec:Sullivan}, we evaluate the Sullivan-process contribution
to the leading-neutron structure function. Figure~\ref{fig:h1f2ln}
compares the H1 measurements of $F_2^{\mathrm{LN}(3)}(x_B,Q^2,x_L)$
with calculations based on the JAM21, xFitter, and GRV pion PDFs. The
results are organized into seven $Q^2$ intervals spanning
$6<Q^2<100~\mathrm{GeV}^2$ and seven $x_L$ intervals covering
$0.32<x_L<0.95$. To better illustrate the $x_\pi$ region probed by the
measurements, $F_2^{\mathrm{LN}(3)}$ is shown as a function of
$x_\pi$, obtained from $x_B$ using Eq.~(\ref{eq:xpixB}), for each
$(Q^2,x_L)$ interval.

Owing to the factorized form of the pion flux $f_{\pi N}(x_L)$ and
pion structure function $F_2^\pi(x_\pi,Q^2)$ in
Eq.~(\ref{eq:F2N_OPE}), the predicted $F_2^{\mathrm{LN}(3)}(x_\pi)$
distributions in each $(Q^2,x_L)$ interval closely follow the behavior
of $F_2^\pi(x_\pi,Q^2)$ shown in Fig.~\ref{fig:f2sf}. For a given
$x_L$ interval, the small-$x_\pi$ rise becomes more pronounced with
increasing $Q^2$, reflecting by DGLAP QCD evolution. Conversely, at
fixed $Q^2$, the distributions retain similar shapes as functions of
$x_\pi$, determined primarily by $F_2^\pi(x_\pi,Q^2)$, while their
overall normalizations vary with $x_L$ through the pion flux factor
$f_{\pi N}(x_L)$ shown in the left panel of Fig.~\ref{fig:pionflux}.

The calculations of the one-pion-exchange contribution based on the
JAM21 pion PDFs show good agreement with the H1 data as functions of
$x_\pi$ across all $Q^2$ bins for the four largest $x_L$ values,
$x_L=0.91$, 0.81, 0.72, and 0.64. The agreement deteriorates for
$x_L<0.6$, where contributions beyond the simple pion-exchange
mechanism, including DIS target fragmentation, are expected to become
increasingly important. As also seen in the figure, the measured
points extend to smaller $x_\pi$ as $x_L$ decreases. This suggests
that incorporating lower-$x_L$ data into future global analyses could
significantly improve constraints on pion PDFs at smaller $x_\pi$,
provided that contributions beyond the OPE mechanism can be reliably
controlled.

To quantify the sensitivity to the pion-flux parameterization,
Fig.~\ref{fig:h1datatotheoryfive} shows the ratios of the H1 data to
the Sullivan-process calculations obtained with the JAM21 pion PDFs
for the five flux models defined in Eq.~(\ref{eq:regulators}),
focusing on the three largest $x_L$ bins, $x_L=0.91$, 0.81, and
0.72. The corresponding dependence on the choice of pion PDF,
evaluated using the $s$-dependent exponential flux, is shown in
Fig.~\ref{fig:h1threedata} of the
Appendix~\ref{sec:appendix_ratios}. In these kinematic regions, the
ratios generally remain close to unity, indicating overall consistency
between the measurements and the theoretical calculations.

Because the JAM21 pion PDFs were determined using the H1
leading-neutron data within the $s$-dependent exponential flux
framework, particularly good agreement with this flux parameterization
is expected. The JAM21~\cite{Cao:2021aci} and
JAM18~\cite{Barry:2018ort} global analyses also introduced overall
normalization factors of 1.26 and 1.17, respectively, for the H1 data
to account for correlated normalization uncertainties. No such
additional normalization factors are applied in the present
calculation. Small differences between our results and those obtained
in the corresponding JAM analyses are therefore
expected. Nevertheless, the other four pion-flux parameterizations
also provide reasonable descriptions of the data, although with
somewhat larger systematic deviations from unity.


\begin{figure*}[!htb]
\centering
\includegraphics[width=\textwidth]{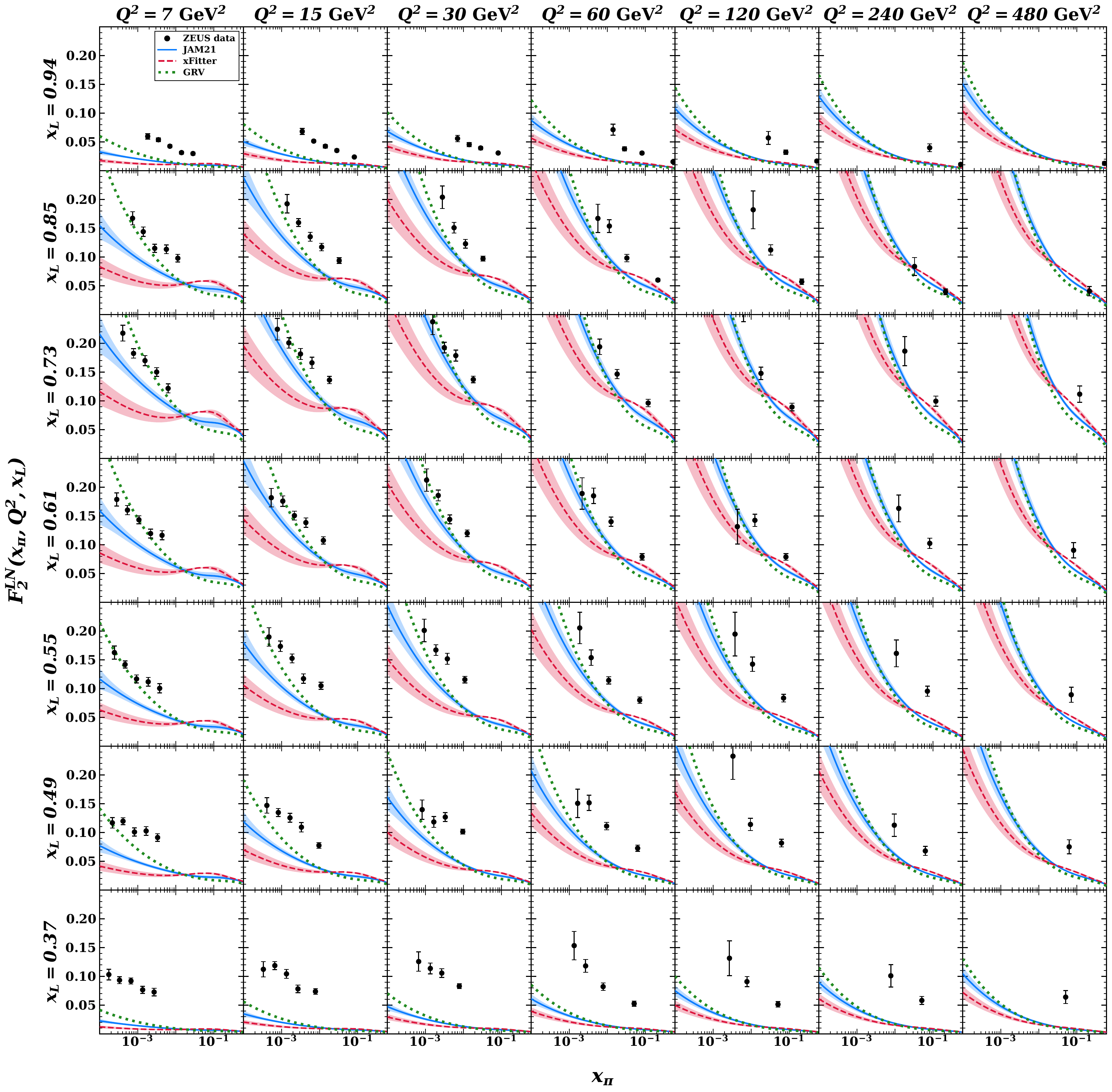}
\caption{The leading-neutron structure function
  $F_2^{\text{LN}(3)}(x_\pi, Q^2, x_L)$ as a function of $x_\pi$ for
  various $x_L$ and $Q^2$ values corresponding to the ZEUS
  kinematics~\cite{ZEUS:2002gig}, computed using the $s$-dependent
  exponential pion flux. Each row represents a fixed value of $x_L$,
  while each column corresponds to a fixed value of $Q^2$. The
  experimental structure function points are reconstructed from the
  measured ratio $r$ via Eq.~(\ref{eq:r_def}) using the CT18NLO proton
  structure function $F_2^p(x_B, Q^2)$~\cite{Hou:2019efy}. The shaded
  bands represent the propagated PDF uncertainties.}
\label{fig:zeusf2ln}
\end{figure*}

\begin{figure*}[!htb]
\centering
\includegraphics[width=\textwidth]{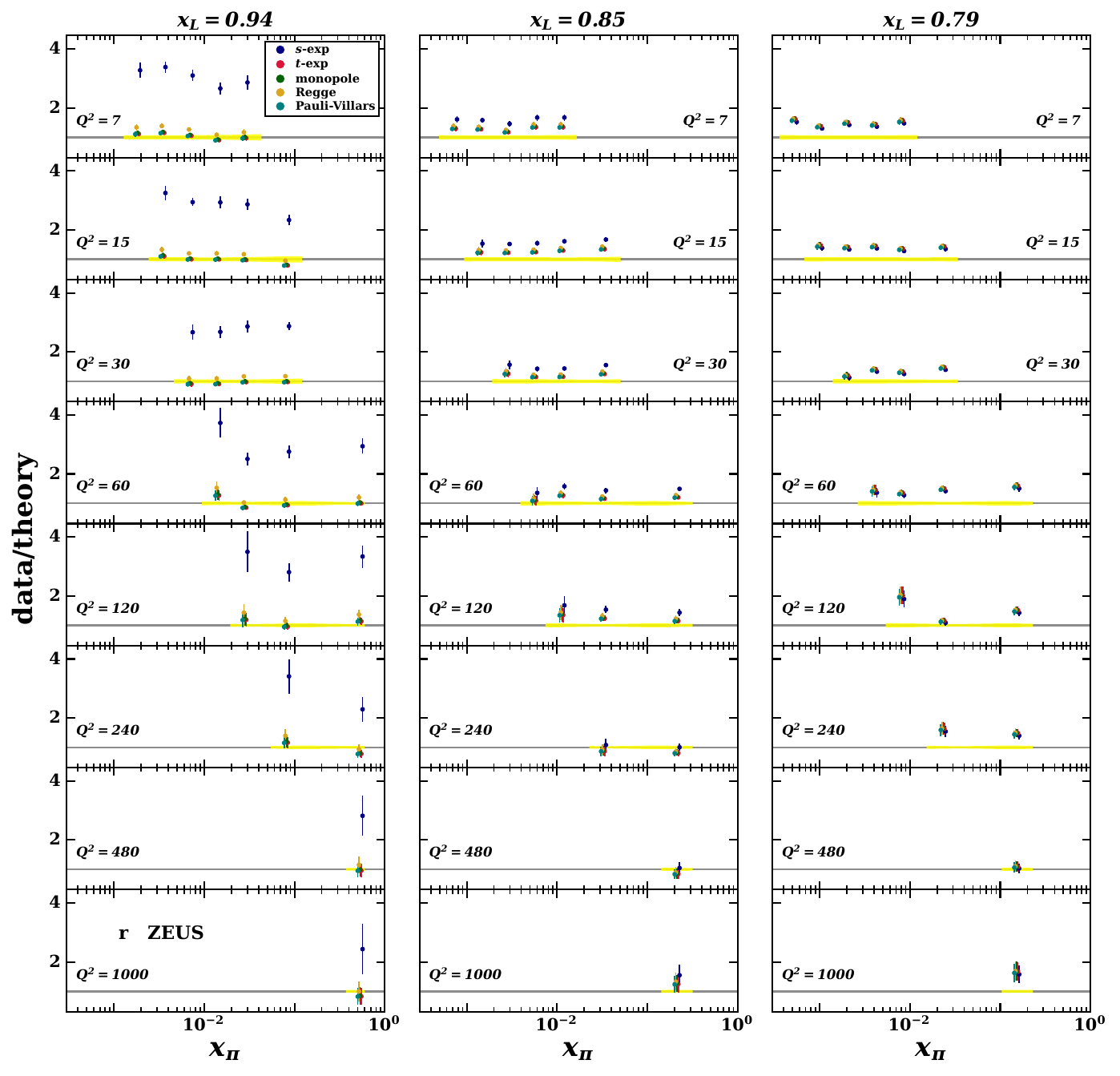}
\caption{Data-to-theory ratios for the leading-neutron structure
  function $F_2^{\mathrm{LN}(3)}$ as functions of $x_\pi$ in selected
  $Q^2$ bins for the three largest ZEUS $x_L$ values, $x_L=0.94$,
  0.85, and 0.79~\cite{ZEUS:2002gig}. The theoretical calculations use
  the JAM21 pion PDF replicas~\cite{Barry:2021osv} together with the
  five pion-flux regulator parameterizations defined in
  Eq.~(\ref{eq:regulators}). The proton structure function
  $F_2^p(x_B,Q^2)$ is evaluated using CT18NLO~\cite{Hou:2019efy}. The
  shaded bands represent the corresponding $1\sigma$ uncertainties
  propagated from the JAM21 PDF replicas.}
\label{fig:zeusfivefluxthreepdf}
\end{figure*}

For comparison with the ZEUS measurements, we convert the measured
ratios $r$ into the leading-neutron structure function
$F_2^{\mathrm{LN}(3)}(x_\pi,Q^2,x_L)$ following Eq.~(\ref{eq:r_SF}),
by dividing by the bin width $\Delta x_L$ and multiplying by the
proton structure function $F_2^p(x_B,Q^2)$ evaluated using the CT18NLO
parameterization~\cite{Hou:2019efy}. We verify that the reconstructed
$F_2^{\mathrm{LN}(3)}$ is negligibly sensitive to the choice of proton
PDF, indicating that the associated proton-PDF uncertainty is small
compared with uncertainties from the pion sector. Following the
procedure used for the H1 data, we compare the ZEUS measurements of
$F_2^{\mathrm{LN}(3)}(x_\pi,Q^2,x_L)$ with the Sullivan-process
predictions in Figs.~\ref{fig:zeusf2ln} and
\ref{fig:zeusfivefluxthreepdf}, using the $s$-dependent exponential
flux and the five pion-flux parameterizations, respectively. For
clarity, only 7 of the 12 available $x_L$ bins and 7 of the 8
available $Q^2$ bins are displayed in Fig.~\ref{fig:zeusf2ln}, while
the complete set of kinematic bins is retained in the numerical
analysis. The dependence of the ZEUS data-to-theory ratios on the
choice of pion PDF is shown in Fig.~\ref{fig:zeusthreedata} of the
Appendix~\ref{sec:appendix_ratios}.

Using the same $s$-dependent exponential pion flux and cutoff
parameter as adopted in the JAM18 analysis, the predicted
Sullivan-process contribution shows noticeably poorer agreement with
the ZEUS measurements at large $x_L$ than with the corresponding H1
data. Most notably, at $x_L=0.94$, the $s$-dependent exponential flux
underestimates the ZEUS $F_2^{\mathrm{LN}(3)}$ measurements by as much
as a factor of 3, as shown in the left panel of
Fig.~\ref{fig:zeusfivefluxthreepdf}. In contrast, the other four
pion-flux parameterizations provide a substantially better description
of the ZEUS data, with data-to-theory ratios remaining closer to unity
in the large-$x_L$ region.

The strong suppression of the Sullivan-process contribution at
$x_L=0.94$ is associated with the rapid decrease of the $s$-dependent
exponential pion flux as $x_L\to1$, as seen in
Fig.~\ref{fig:pionflux}. Within the present implementation, the
combination of this flux parameterization with the JAM pion PDFs
therefore does not provide a simultaneous description of the H1 and
ZEUS measurements at large $x_L$. This observation appears to differ
from the results reported in Refs.~\cite{Barry:2018ort,Barry:2021osv},
where the H1 and ZEUS leading-neutron data at $x_L>0.8$ were described
in global analyses employing the same form of the $s$-dependent
exponential regulator. The JAM21~\cite{Barry:2021osv} and
JAM18~\cite{Barry:2018ort} analyses additionally allowed overall
normalization factors of $0.95$ and $0.964$, respectively, for the
ZEUS data. These few-percent normalization shifts are, however, far
too small to account for the factor-of-3 discrepancy observed at
$x_L=0.94$.

\section{DIS target fragmentation simulated by \textsc{PYTHIA}}
\label{sec:DIS_fragmentation}

As established in Sec.~\ref{sec:Data_Sullivan}, the OPE Sullivan
mechanism provides a good description of the leading-neutron structure
function primarily in the forward region, $x_L\gtrsim0.6$. At smaller
longitudinal momentum fractions, $x_L<0.6$, the pion-exchange
contribution decreases rapidly and systematically undershoots the
measured cross section, indicating the increasing importance of
additional production mechanisms. In this study, we focus on DIS
target fragmentation as the principal non-OPE contribution at low
$x_L$, motivated by the phenomenological interpretations adopted in
the H1 and ZEUS analyses. The fragmentation contribution is simulated
using \textsc{Pythia}~8~\cite{Bierlich:2022pfr}. For comparison, the
original H1 analysis modeled the hadronic final state using
\textsc{Ariadne} together with the Lund string model implemented in
\textsc{Jetset} within the \textsc{Django}
framework~\cite{H1:2010hym}, while the ZEUS analysis employed standard
DIS Monte Carlo generators to estimate non-OPE
backgrounds~\cite{ZEUS:2002gig}.

In \textsc{Pythia}, target fragmentation emerges from the combined
treatment of the beam remnant, color connections to the
hard-scattering system, parton showers, and subsequent hadronization
through the Lund string model. The beam-remnant and color-reconnection
dynamics are controlled through the \texttt{BeamRemnants} and
\texttt{ColourReconnection}~\cite{Pythia8317Manual} settings. We
compare the default beam-remnant configuration,
\texttt{BeamRemnants:remnantMode = 0}, with the updated option,
\texttt{BeamRemnants:remnantMode = 1}, using the QCD-based
color-reconnection model, \texttt{ColourReconnection:mode = 1}, in
both cases. To assess the sensitivity to the color-reconnection
prescription, we also consider the gluon-move model,
\texttt{ColourReconnection:mode = 2}, which rearranges soft gluons
among string systems to reduce the total string length prior to
hadronization.

In addition, we examine the forward-physics tune developed in
Ref.~\cite{Fieg:2023kld}. Although this tune was originally
constrained by LHCf measurements of forward neutron, pion, and photon
production in proton--proton collisions, its modified beam-remnant
parameters directly affect the fragmentation of the proton remnant and
therefore provide a useful test of model variations in DIS target
fragmentation. Finally, we investigate the dependence of the
fragmentation contribution on the assumed proton partonic
structure. Predictions obtained with the CT18NLO proton PDF
set~\cite{Hou:2019efy}, accessed through
LHAPDF~\cite{Buckley:2014ana}, are compared with those based on the
default NNPDF2.3 LO set in \textsc{Pythia}. All other generator
settings are held fixed unless stated otherwise. The \textsc{Pythia}
configurations considered in this study are summarized in
Table~\ref{tab:pythia}.

\begin{table}[htbp]
\caption{\textsc{Pythia}~8 configurations considered in this work and
  the corresponding modified parameters.}
\label{tab:pythia}
\centering
\renewcommand{\arraystretch}{1.2}
\setlength{\tabcolsep}{3pt}
\footnotesize
\begin{tabular}{ll}
\hline\hline
\textbf{Configuration} & \textbf{Modified \textsc{Pythia}~8 Settings} \\
\hline
\textsc{Pythia} Default & Appendix~\ref{sec:appendix_pythia} \\
\hline
CT18 NLO & \texttt{PDF:pSet = LHAPDF6:CT18NLO} \\
\hline
CT18 NLO + Remnant & \texttt{PDF:pSet = LHAPDF6:CT18NLO} \\
 & \texttt{BeamRemnants:remnantMode = 1} \\
 & \texttt{ColourReconnection:mode = 1} \\
\hline
CT18 NLO + Tuning~\cite{Fieg:2023kld} & \texttt{PDF:pSet = LHAPDF6:CT18NLO} \\
 & \texttt{ColourReconnection:} \\
 & \texttt{~~allowJunctions = on} \\
 & \texttt{BeamRemnants:dampPopcorn = 0} \\
 & \texttt{BeamRemnants:} \\
 & \texttt{~~hardRemnantBaryon = on} \\
 & \texttt{BeamRemnants:aRemnantBaryon = 0.36} \\
 & \texttt{BeamRemnants:bRemnantBaryon = 1.69} \\
 & \texttt{BeamRemnants:} \\
 & \texttt{~~primordialKTsoft = 0.58} \\
 & \texttt{BeamRemnants:} \\
 & \texttt{~~primordialKTremnant = 0.58} \\
\hline
CT18 NLO + Gluon & \texttt{PDF:pSet = LHAPDF6:CT18NLO} \\
 & \texttt{BeamRemnants:remnantMode = 0} \\
 & \texttt{ColourReconnection:mode = 2} \\
\hline\hline
\end{tabular}
\end{table}

Using the \textsc{Pythia} configurations described above, we generate
inclusive DIS event samples and select events containing at least one
final-state neutron that satisfy the experimental acceptance cuts
summarized in Table~\ref{tab:kinematics}. The semi-inclusive
differential cross section, $d^{3}\sigma(ep\to e'nX)/(dx_B dQ^{2}
dx_L)$, is obtained from the corresponding event yields normalized by
the total number of generated events, $N_{\mathrm{gen}}$, and the
generated inclusive DIS cross section, $\sigma_{\mathrm{gen}}$,
provided by \texttt{pythia.info.sigmaGen()}. To reduce statistical
fluctuations in the determination of $\sigma_{\mathrm{gen}}$,
$5\times10^{8}$ events are generated for each configuration, for which
the resulting cross section is found to be numerically stable.

\begin{figure*}[!htbp]
\centering
\begin{minipage}{0.49\textwidth}
    \centering
    \includegraphics[width=\textwidth]{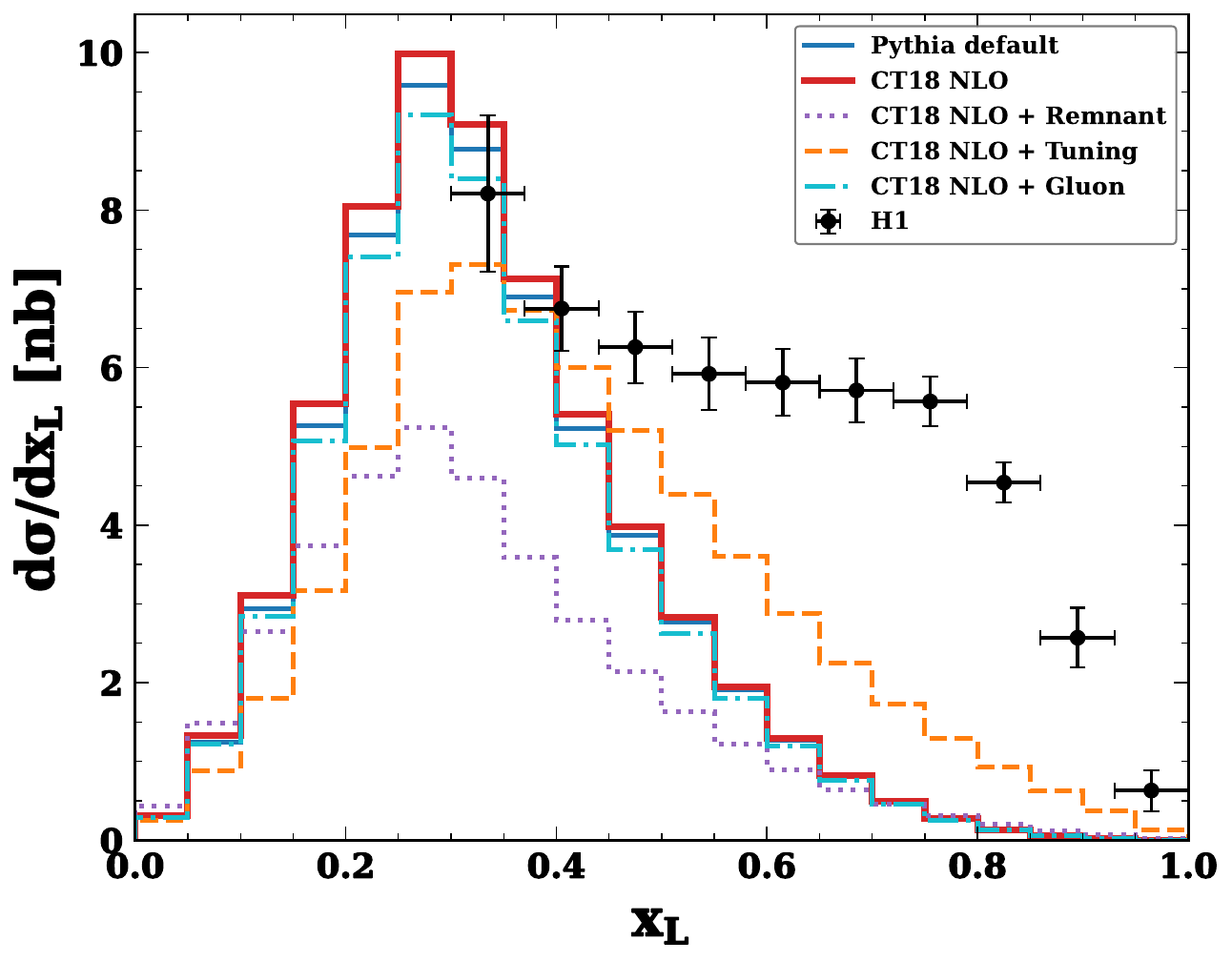}
\end{minipage}
\hfill
\begin{minipage}{0.49\textwidth}
    \centering
    \includegraphics[width=\textwidth]{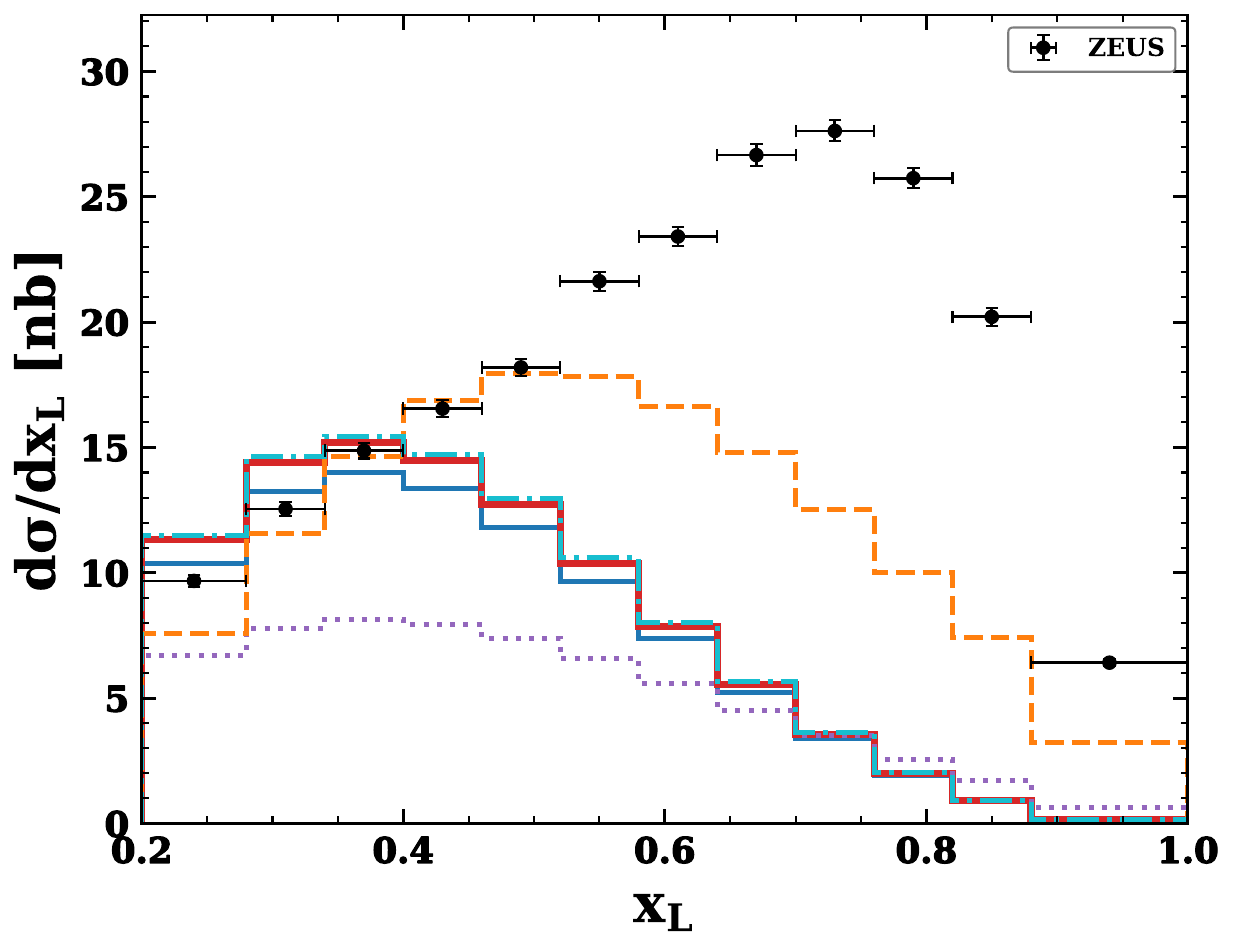}
\end{minipage}
\caption{Differential cross section $d\sigma/dx_L$ for leading-neutron
  production, calculated using the \textsc{Pythia}~8 configurations
  listed in Table~\ref{tab:pythia}. The left and right panels
  correspond to the H1 phase space, with $p_T<0.2~\mathrm{GeV}$, and
  the ZEUS phase space, with $p_T \leq 0.656 x_L~\mathrm{GeV}$,
  respectively.}
\label{fig:dsigma_combined}
\end{figure*}

Figure~\ref{fig:dsigma_combined} compares the DIS leading-neutron
differential cross section $d\sigma/dx_L$ predicted by the different
\textsc{Pythia} configurations with the H1 and ZEUS measurements under
the acceptance cuts listed in Table~\ref{tab:kinematics}. In both
phase spaces, the DIS target-fragmentation contribution is largest at
low and intermediate $x_L$ and decreases rapidly toward larger
$x_L$. The predicted fragmentation contribution accounts for a
substantial fraction of the measured cross section for
$x_L\lesssim0.5$, but systematically undershoots the data in the
forward region, $x_L\gtrsim0.6$. This behavior supports the increasing
importance of the OPE Sullivan contribution at large $x_L$, consistent
with the use of this region in pion PDF extractions.

A comparison of the generator configurations shows that the baseline
``\textsc{Pythia} default'', ``CT18 NLO'', and ``CT18 NLO + Gluon''
predictions yield nearly identical $d\sigma/dx_L$ distributions. This
indicates that, within the configurations considered, the modeled
fragmentation contribution is only weakly sensitive to the choice of
proton PDF and to the gluon-move color-reconnection prescription. In
contrast, the ``CT18 NLO + Remnant'' configuration suppresses the
cross section over the full $x_L$ range, demonstrating a stronger
sensitivity to the treatment of the beam remnant. The ``CT18 NLO +
Tuning'' configuration produces a substantially harder neutron
spectrum, with the largest enhancement for $x_L\gtrsim0.5$. This
behavior is associated with the modified beam-remnant parameters of
the forward-physics tune, including suppression of the popcorn
mechanism, modifications to the remnant-baryon fragmentation function,
and a reduced primordial transverse momentum, which collectively favor
the production of energetic forward baryons.

\section{Comparison of H1 and ZEUS data with combined Sullivan-process and DIS target-fragmentation contributions}
\label{sec:Data_CombinedResults}

For a complete description of the H1 and ZEUS data sets, both the OPE
Sullivan mechanism and the DIS target-fragmentation contribution are
taken into account. In the H1 analysis~\cite{H1:2010hym}, the
leading-neutron data were described by superimposing
target-fragmentation events generated with \textsc{Django} on the OPE
contribution simulated with \textsc{Rapgap-$\pi$}, with empirical
normalization factors of $1.2$ and $0.65$ applied to the respective
components~\cite{H1:2010hym}. In the present study, no additional ad
hoc normalization factors are introduced. This allows us to test
whether an unscaled combination of the Sullivan-process contribution
and \textsc{Pythia} DIS target fragmentation can provide a consistent
description of the data across the measured phase space.

We compare the resulting predictions with the measured three-fold
leading-neutron structure function
$F_{2}^{\mathrm{LN}(3)}(x_B,Q^2,x_L)$ as a function of $x_L$ in bins
of $(Q^2,x_B)$. The phase space is partitioned according to the
binning definitions used by the H1 and ZEUS Collaborations. For each
bin, the Sullivan-process contribution is evaluated at the bin-center
values of $Q^2$ and $x_B$, while the \textsc{Pythia}
target-fragmentation contribution is obtained by integrating over
generated events that satisfy the corresponding bin boundaries. The
resulting differential cross section $d\sigma/dx_L$ is converted to
$F_2^{\mathrm{LN}(3)}$ using Eq.~(\ref{eq:ln_cross_section}), with the
inelasticity evaluated as $y=Q^2/(x_Bs)$. Center-of-mass energies of
$\sqrt{s}=318~\mathrm{GeV}$ for H1 and $\sqrt{s}=300~\mathrm{GeV}$ for
ZEUS are used to reproduce the respective collider conditions.

\begin{figure*}[!htbp]
\centering
\includegraphics[width=\linewidth]{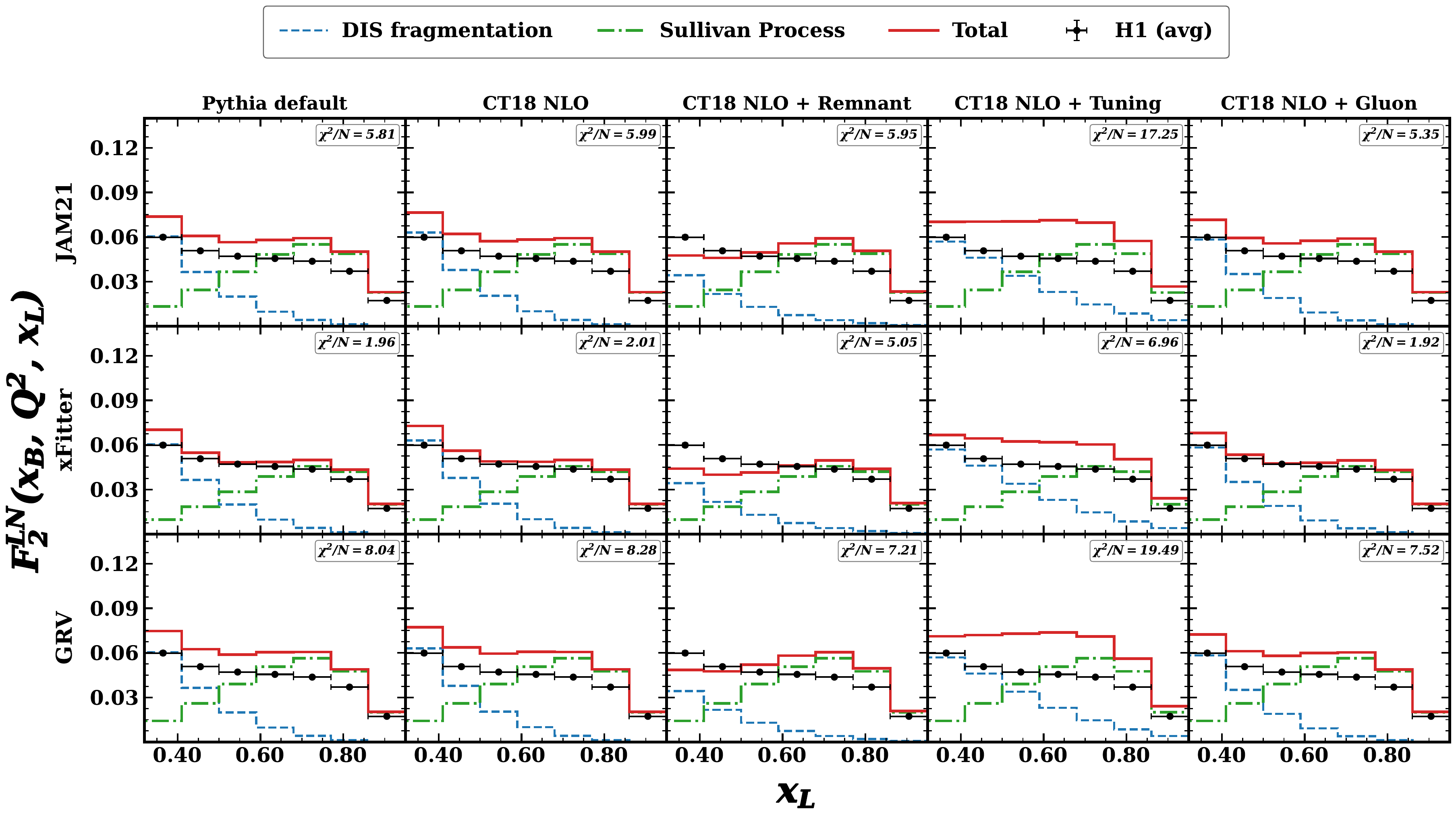}
\includegraphics[width=\linewidth]{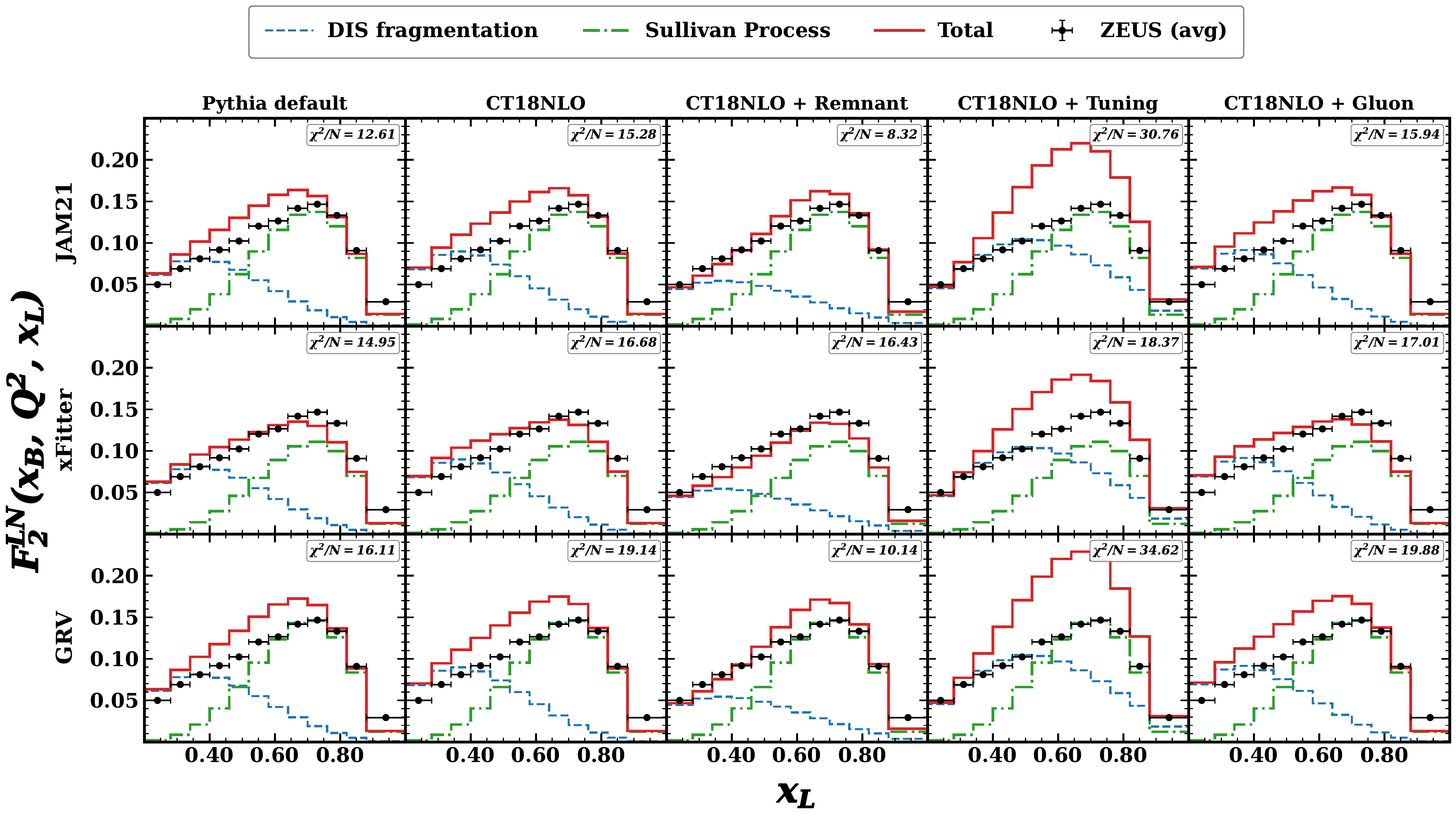}
\caption{Predicted leading-neutron structure function
  $F_{2}^{\mathrm{LN}(3)}(x_L)$ compared with the H1 (top) and ZEUS
  (bottom) measurements averaged over the corresponding $x_B$ and
  $Q^2$ bins. The rows correspond to the JAM21, xFitter, and GRV pion
  PDF sets used to calculate the OPE Sullivan contribution, while the
  columns correspond to the five \textsc{Pythia}~8
  target-fragmentation configurations listed in
  Table~\ref{tab:pythia}. The agreement between the data and
  predictions is quantified by $\chi^2/N$ in each panel.}
\label{fig:f2lntotalbackgroundPDF_h1zeus}
\end{figure*}

Figure~\ref{fig:f2lntotalbackgroundPDF_h1zeus} compares the predicted
semi-inclusive leading-neutron structure function
$F_{2}^{\mathrm{LN}(3)}(x_L)$ as a function of $x_L$ with the H1 and
ZEUS measurements averaged over the corresponding $x_B$ and $Q^2$
bins. The predictions combine the OPE Sullivan contribution,
calculated using the JAM21, xFitter, and GRV pion PDF sets, with the
five \textsc{Pythia}~8 target-fragmentation configurations listed in
Table~\ref{tab:pythia}, yielding 15 model combinations arranged in a
$3\times5$ grid. The agreement between the data and predictions is
quantified by $\chi^2/N$ in each panel. Among the configurations
considered, `CT18 NLO'' and `CT18 NLO + Remnant'' provide the best
overall descriptions of the averaged H1 and ZEUS data,
respectively. Since the $d\sigma/dx_L$ distributions obtained with the
``\textsc{Pythia} default'', ``CT18 NLO'', and ``CT18 NLO + Gluon''
configurations, as well as their corresponding $\chi^2/N$ values, are
very similar, we adopt the ``CT18 NLO'' configuration as a
representative choice for the following comparison with the H1 data.

\begin{figure*}[!htbp]
\centering
\includegraphics[width=\textwidth]{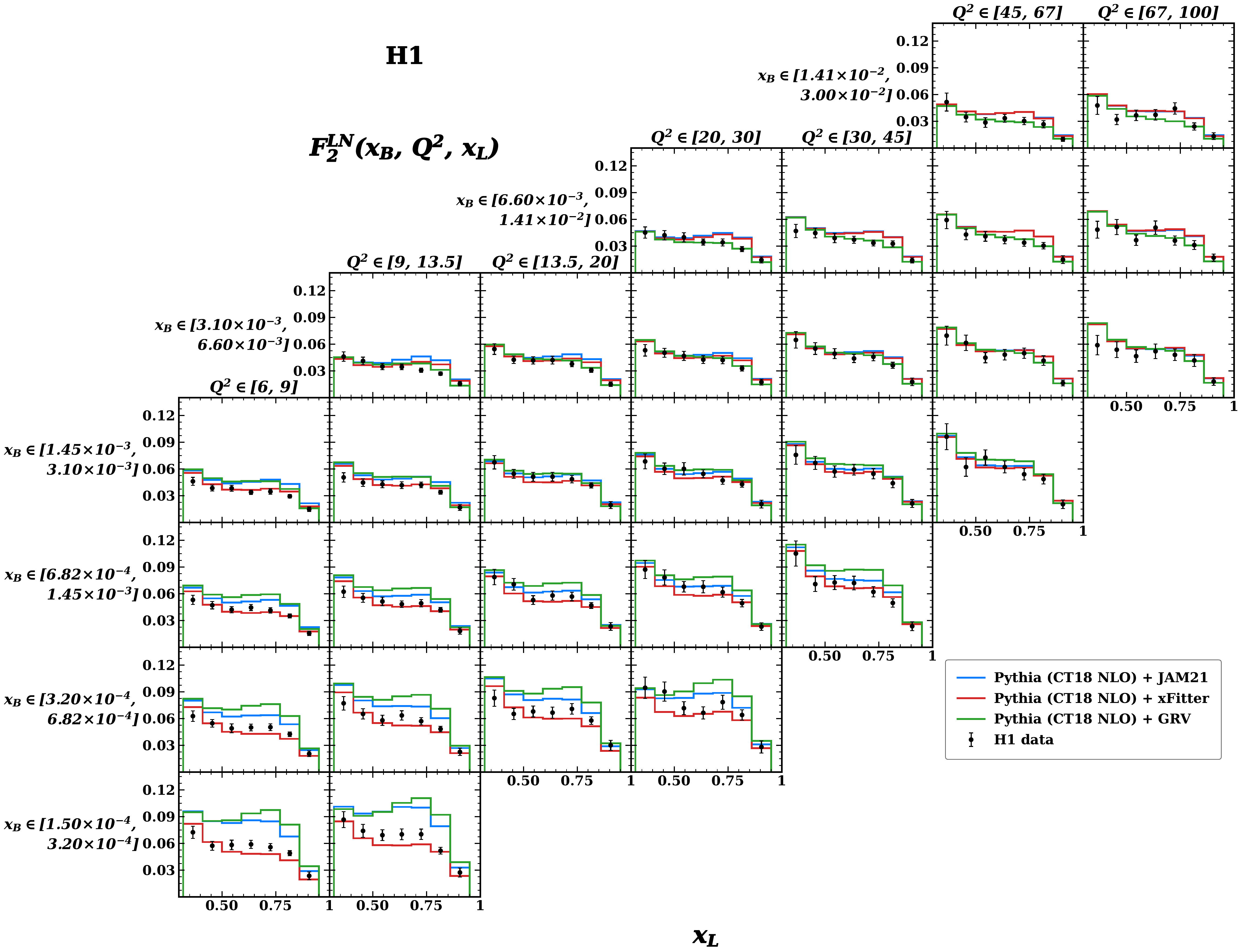}
\caption{Three-fold leading-neutron structure function
  $F_{2}^{\mathrm{LN}(3)}(x_B,Q^2,x_L)$ as a function of $x_L$ for the
  H1 data~\cite{H1:2010hym} with the requirement
  $p_T<0.2~\mathrm{GeV}$. The curves show the sum of the
  \textsc{Pythia} DIS target-fragmentation contribution obtained with
  the ``CT18 NLO'' configuration and the Sullivan-process
  contributions calculated using the JAM21, xFitter, and GRV pion
  PDFs. Each column corresponds to a fixed $Q^2$ bin and each row to a
  fixed $x_B$ bin.}
\label{fig:f2lnh1total}
\end{figure*}

\begin{figure*}[!htbp]
\centering
\includegraphics[width=\textwidth]{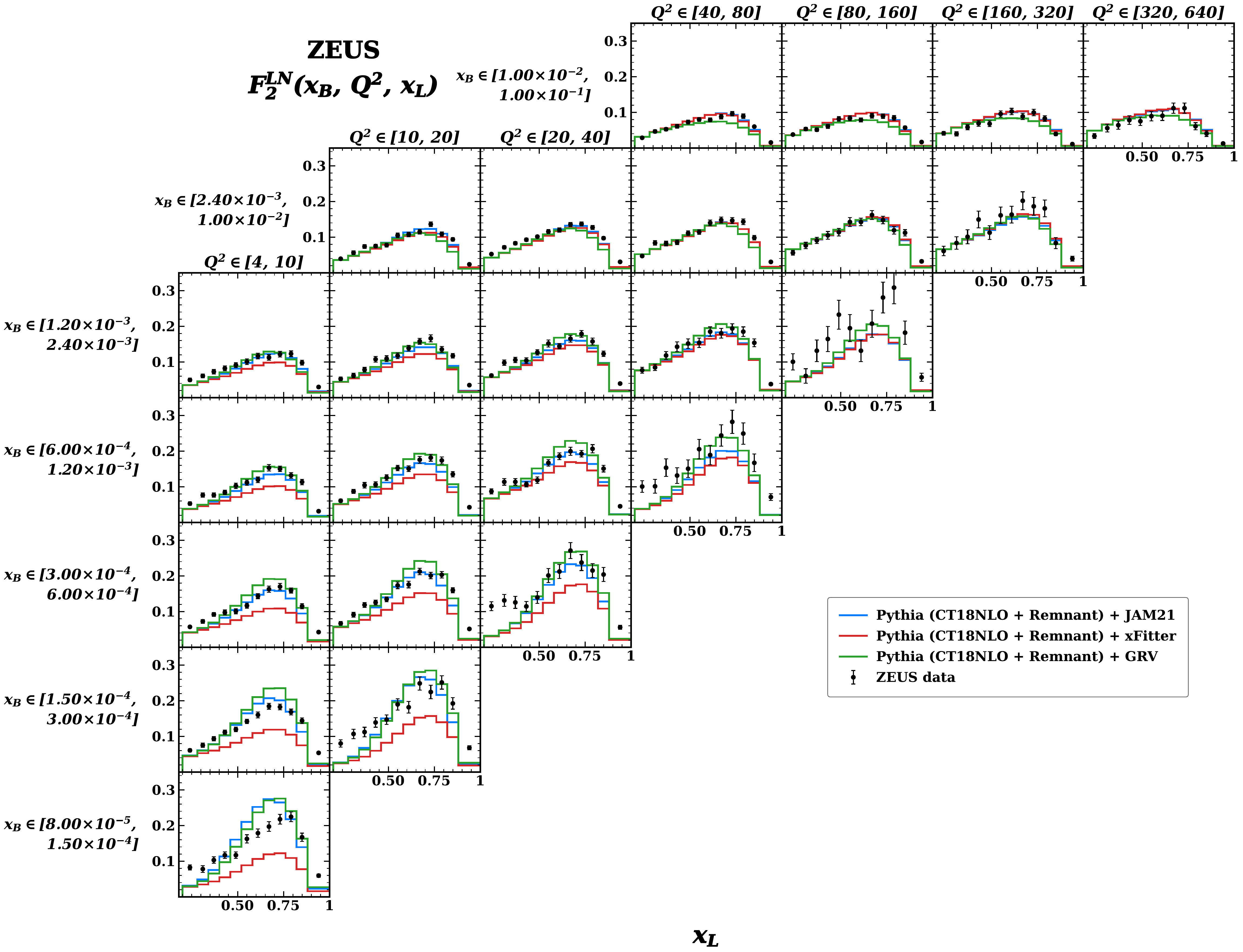}
\caption{Same as Fig.~\ref{fig:f2lnh1total}, but for the ZEUS
  data~\cite{ZEUS:2002gig} with the neutron transverse-momentum
  requirement $p_T \leq 0.656 x_L~\mathrm{GeV}$. The DIS
  target-fragmentation contribution is calculated using the ``CT18
  NLO+Remnant'' configuration in \textsc{Pythia}.}
\label{fig:f2lnzeustotal}
\end{figure*}

As shown in Fig.~\ref{fig:f2lnh1total}, the sum of the
Sullivan-process and DIS target-fragmentation contributions obtained
with the ``CT18 NLO'' configuration reproduces the main features of
the H1 $F_2^{\mathrm{LN}(3)}$ measurements across the full $(Q^2,x_B)$
phase space. Over the intermediate region $3.10\times10^{-3}\lesssim
x_B\lesssim1.41\times10^{-2}$, the predictions based on the JAM21,
xFitter, and GRV pion PDFs exhibit very similar $x_L$ dependence and
closely follow the measured data. More pronounced differences among
the pion PDF sets emerge in the lowest-$x_B$ bins,
$x_B<3.10\times10^{-3}$. Through the kinematic relation
$x_\pi=x_B/(1-x_L)$ [Eq.~(\ref{eq:xpixB})], these differences can be
traced to the distinct small-$x_\pi$ behavior of the pion structure
functions shown in Fig.~\ref{fig:f2sf}. The overall $\chi^2/N$ values
for the full data set are 6.0, 2.0, and 8.3 for the JAM21, xFitter,
and GRV pion PDFs, respectively, with xFitter providing the best
quantitative agreement.

The growing separation among the predictions at low $x_B$ originates
primarily from the Sullivan-process contribution, since the DIS
target-fragmentation component is independent of the pion PDF
choice. Through the relation $x_\pi=x_B/(1-x_L)$, decreasing $x_B$ at
fixed $x_L$ probes progressively smaller values of $x_\pi$, where the
three pion PDF sets exhibit their largest differences. The resulting
spread in the small-$x_\pi$ pion structure is therefore reflected
directly in the predicted $F_2^{\mathrm{LN}(3)}$ at low $x_B$ and
intermediate $x_L$. The smaller $\chi^2/N$ obtained with xFitter
indicates that, within the present framework, its small-$x_\pi$
behavior is more compatible with the low-$x_B$ H1 leading-neutron data
than those of JAM21 and GRV.

The corresponding comparison for ZEUS is shown in
Fig.~\ref{fig:f2lnzeustotal}, using the ``CT18 NLO + Remnant''
fragmentation configuration together with the JAM21, xFitter, and GRV
pion PDF sets. In contrast to the H1 comparison, JAM21 provides the
best overall agreement among the three pion PDF sets, with
$\chi^2/N=8.3$, compared with 16.4 for xFitter and 10.1 for GRV. At
low $x_B$ and intermediate $x_L$, the xFitter and GRV predictions
generally lie below and above the JAM21 result, respectively, while
the JAM21 prediction follows the ZEUS measurements more closely over
much of the displayed phase space.

Overall, the present framework provides a qualitatively reasonable
description of the H1 and ZEUS measurements over the full measured
$x_L$ range, although the relatively large $\chi^2/N$ values indicate
that a quantitatively satisfactory description has not yet been
achieved. Nevertheless, it is encouraging that the main features of
both data sets can be reproduced without introducing additional ad hoc
normalization factors. It should also be emphasized that the
parameters and functional forms governing the pion flux are correlated
with the extracted pion PDFs. An improved quantitative description
could be achieved in future global analyses where the regulator cutoff
parameter $\Lambda$ in Eq.~(\ref{eq:regulators}) is fitted
simultaneously with the pion PDFs and constrained directly by the
data. Additional comparisons illustrating the dependence on the DIS
target-fragmentation model and pion PDF choice are presented in
Appendix~\ref{sec:appendix_F2Ntotal}.

\section{Leading neutron Process at EIC}
\label{sec:EIC}

\begin{figure*}[!hbtp]
\centering
\includegraphics[width=\textwidth]{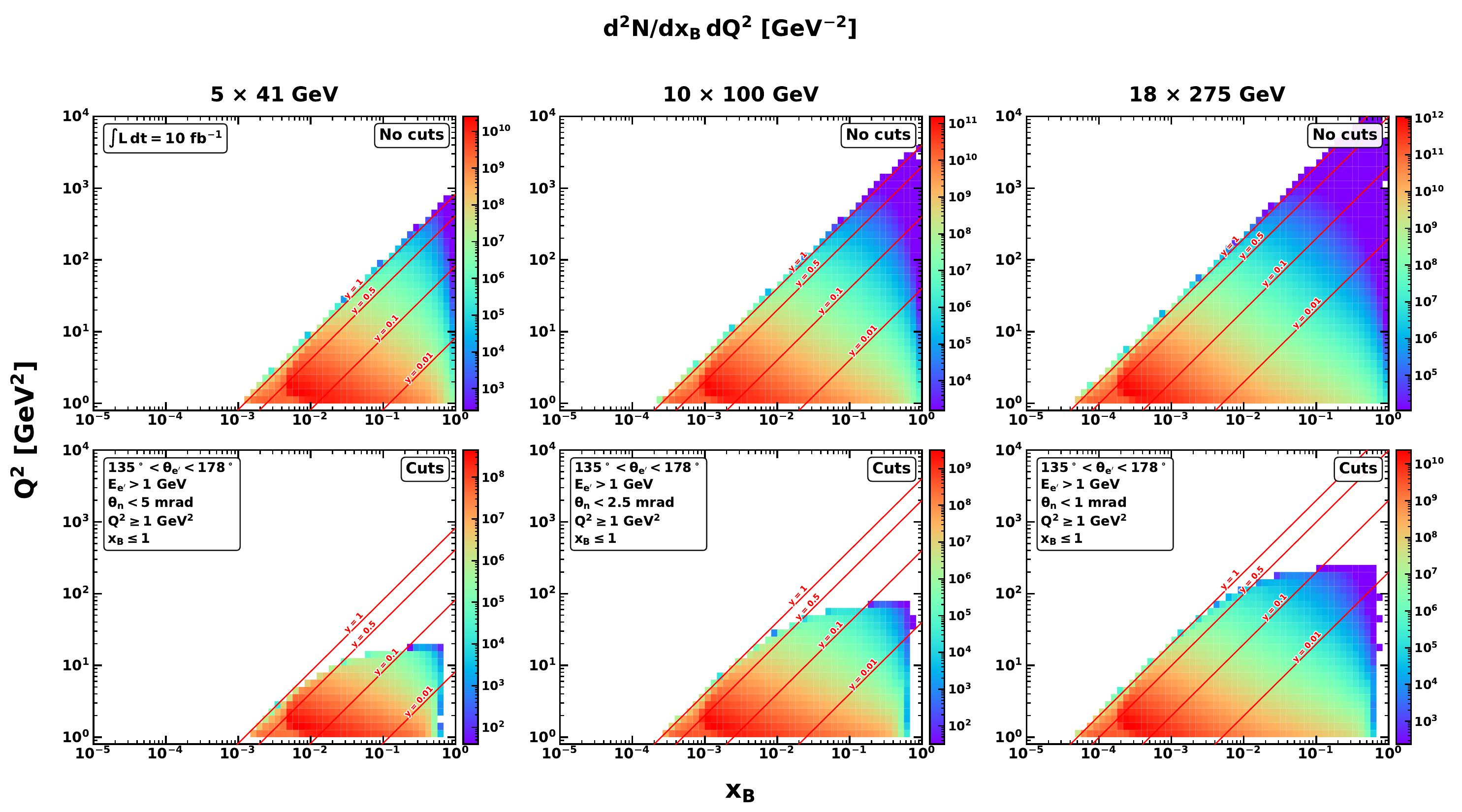}
\caption{Differential leading-neutron event yield
  $d^{2}N/(dx_B,dQ^{2})$ in the $(x_B,Q^{2})$ plane for an integrated
  luminosity of $10~\mathrm{fb}^{-1}$ at the three EIC beam-energy
  configurations, $5\times41$, $10\times100$, and
  $18\times275~\mathrm{GeV}$, shown from left to right. The upper
  panels show the generated-event distributions before detector-level
  selections, while the lower panels show the corresponding yields
  after applying the EIC fiducial requirements described in the text.}
\label{fig:eicphasespace}
\end{figure*}

In this section, we present predictions of differential cross-section
for three beam-energy configurations at the U.S. EIC and identify the
kinematic regions that offer enhanced sensitivity to pion structure
while minimizing contributions from DIS target
fragmentation. Figure~\ref{fig:eicphasespace} shows the differential
DIS leading-neutron event yield $d^2N/(dx_B,dQ^2)$ simulated with
\textsc{Pythia}~8 using CT18NLO for three benchmark EIC $e+p$
beam-energy configurations: $5\times41~\mathrm{GeV}$,
$10\times100~\mathrm{GeV}$, and
$18\times275~\mathrm{GeV}$~\cite{AbdulKhalek:2021gbh}, assuming an
integrated luminosity of $10~\mathrm{fb}^{-1}$. The top panels display
the phase-space distributions of generated events before
detector-level selections, while the bottom panels show the
corresponding yields after applying the core EIC fiducial
requirements~\cite{AbdulKhalek:2021gbh}: scattered-electron energy
$E_e'>1~\mathrm{GeV}$, $Q^2>1~\mathrm{GeV}^2$, electron polar-angle
acceptance $135^\circ<\theta_e'<178^\circ$, and forward-neutron
acceptance $\theta_n<5~\mathrm{mrad}$. Although standard EIC
projections commonly impose $y>0.01$ to ensure reliable reconstruction
at low inelasticity~\cite{AbdulKhalek:2021gbh}, this requirement is
not applied here in order to illustrate the kinematic reach before
imposing an explicit lower-$y$ selection.

The forward-neutron angular acceptance of the EIC ZDC extends well
beyond the $\theta_n<0.8~\mathrm{mrad}$ acceptance of the ZEUS
measurement. Together with the spatial resolution of the ZDC, this
broader angular coverage provides sensitivity to the neutron
transverse momentum and hence to the momentum transfer $t$, enabling
differential studies of the $t$ dependence of leading-neutron
production.

The generated-event distributions exhibit the strong $Q^2$ dependence
characteristic of the DIS cross sections, expanding toward smaller
$x_B$ values as the center-of-mass energy increases due to the
relation $Q^2 = s\,x_B\,y$. Imposing the fiducial cuts severely
reduces the phase space for the $5\times41~\text{GeV}$ setting, where
the reduced $\sqrt{s}$ compresses the region between the $Q^2 >
1~\text{GeV}^2$ cut and the physical boundary $y = 1$. In contrast,
the highest-energy $18\times275~\text{GeV}$ configuration provides the
broadest phase-space coverage, reaching down to $x_B \sim
10^{-4}$. This high-energy setting exposes the small-$x_\pi$ regime
dominated by sea-quark and gluon of the pion. Meanwhile, the
lower-energy settings provide enhanced statistics at large $x_\pi$,
offering a critical overlap region to cross-check pion PDFs extracted
from leading-neutron DIS against those derived from fixed-target
Drell-Yan experiments.

\begin{figure}[!htbp]
\centering \includegraphics[width=\linewidth]{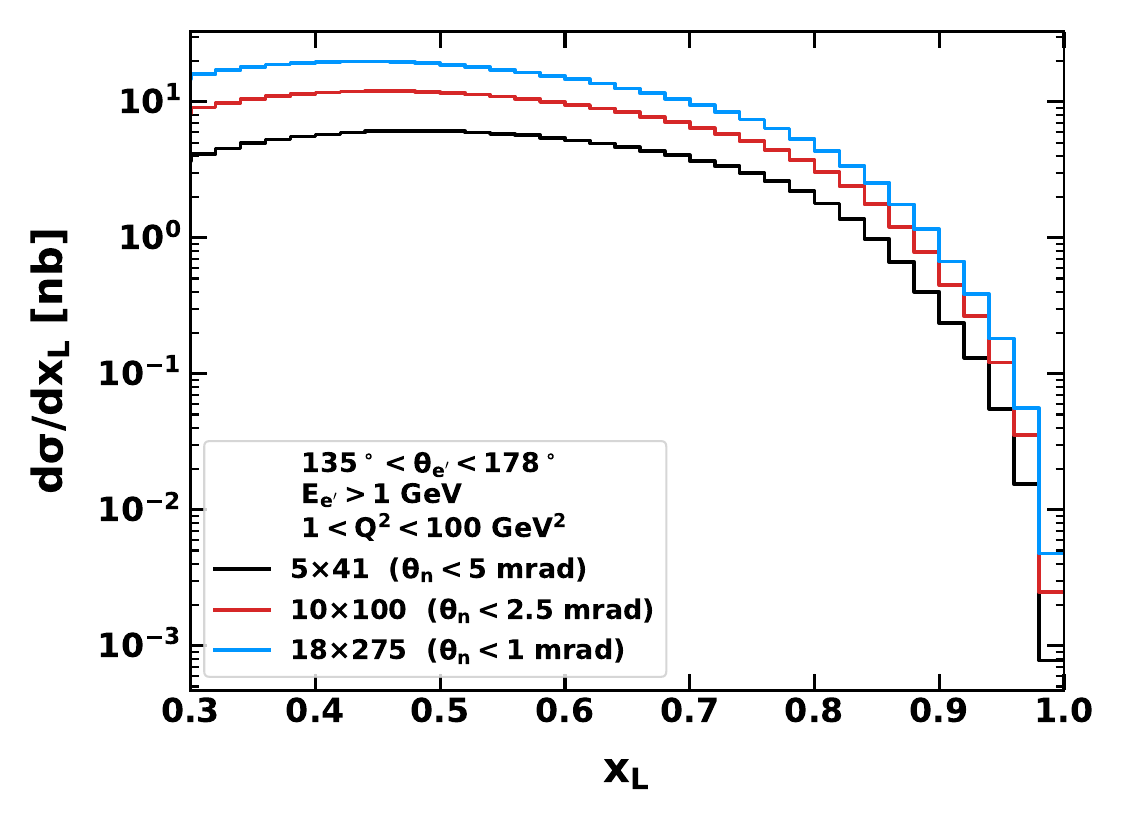}
\caption{The leading-neutron differential cross section
  $d\sigma/dx_{L}$ versus $x_{L}$ for the $5\times41$, $10\times100$,
  and $18\times275~\text{GeV}$ EIC beam-energy settings under the
  fiducial cuts specified in the text.}
\label{fig:dsigmadxleic}
\end{figure}

Figure~\ref{fig:dsigmadxleic} shows the \textsc{Pythia} prediction for
the DIS target-fragmentation contribution to the leading-neutron
differential cross section $d\sigma/dx_L$ for the three nominal EIC
beam-energy configurations. We find that applying the same
forward-neutron angular acceptance to all three configurations leads
to substantially larger target-fragmentation contributions at higher
proton-beam energies, because a fixed angular cut corresponds to a
larger allowed neutron transverse momentum. To suppress this
contribution and retain a sizable relative Sullivan-process component,
we therefore impose progressively tighter neutron angular requirements
at higher energies: $\theta_n<1$, 2.5, and 5~mrad for the high-,
intermediate-, and low-energy configurations, respectively. The
predicted differential yield generally increases with center-of-mass
energy, reflecting the broader accessible DIS phase space at larger
$\sqrt{s}$. Across all three configurations, the target-fragmentation
spectrum remains relatively broad up to $x_L\lesssim0.7$ and decreases
rapidly in the region $x_L\sim0.8$--0.9. This suppression at large
$x_L$ is favorable for Sullivan-process measurements because the
fragmentation contribution becomes comparatively small in the
kinematic region where pion exchange is expected to be most
important. The large-$x_L$ region therefore provides a favorable
window for studying the Sullivan process and constraining pion
structure, particularly when combined with appropriately restrictive
forward-neutron angular cuts.

\begin{figure*}[t]
\centering
\includegraphics[width=\textwidth]{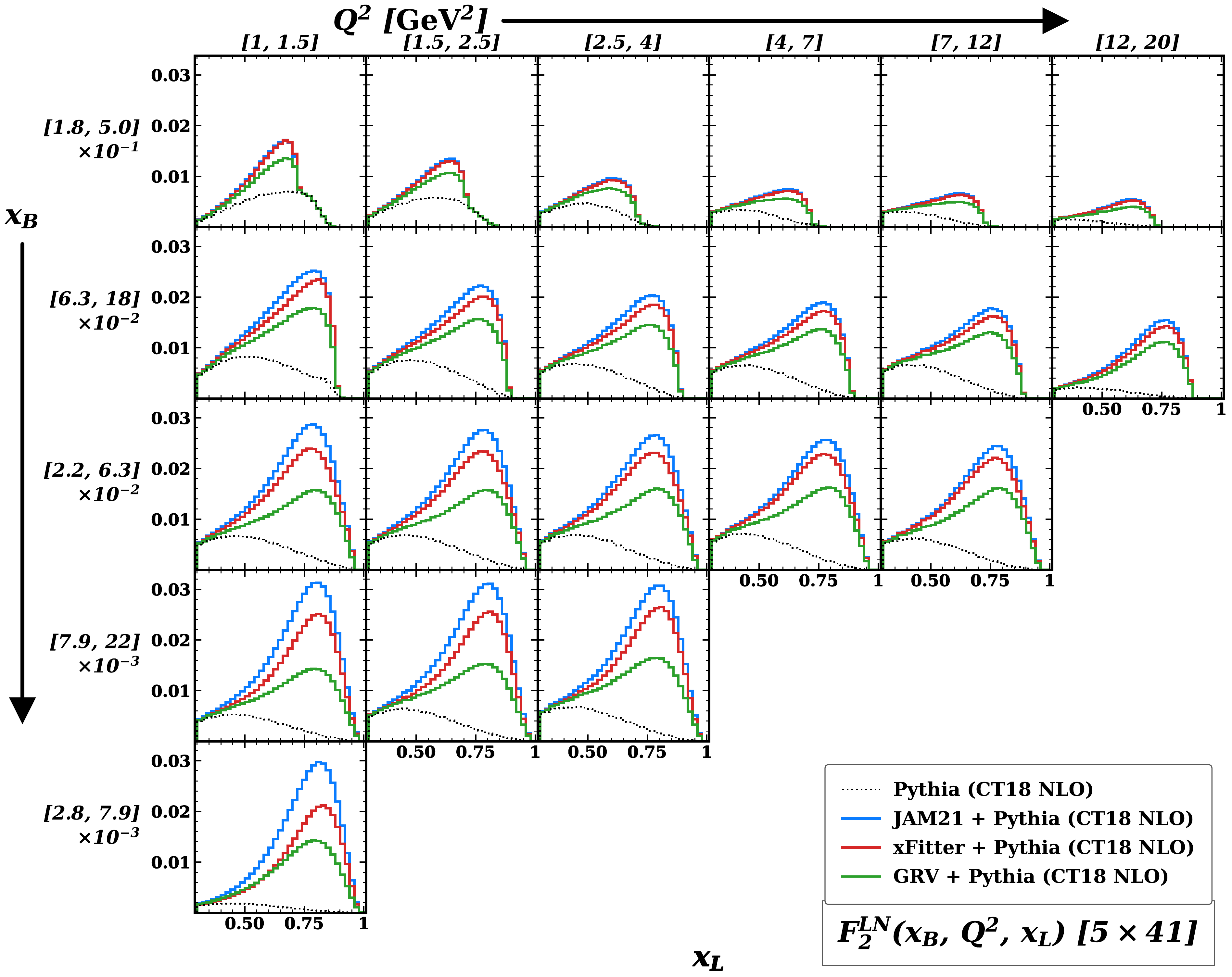}
\caption{Three-fold leading-neutron structure function
  $F_2^{\mathrm{LN}(3)}(x_B,Q^2,x_L)$ as a function of $x_L$ for the
  $5\times41~\mathrm{GeV}$ EIC configuration in the $(x_B,Q^2)$ bins
  defined in Table~\ref{tab:eic}. Each panel shows the
  \textsc{Pythia}~8 DIS target-fragmentation contribution obtained
  with CT18NLO, together with the total predictions formed by adding
  the OPE contribution calculated using the JAM21, xFitter, and GRV
  pion PDFs with the $s$-dependent exponential flux.}
\label{fig:eic541}
\end{figure*}

\begin{figure*}[t]
\centering
\includegraphics[width=\textwidth]{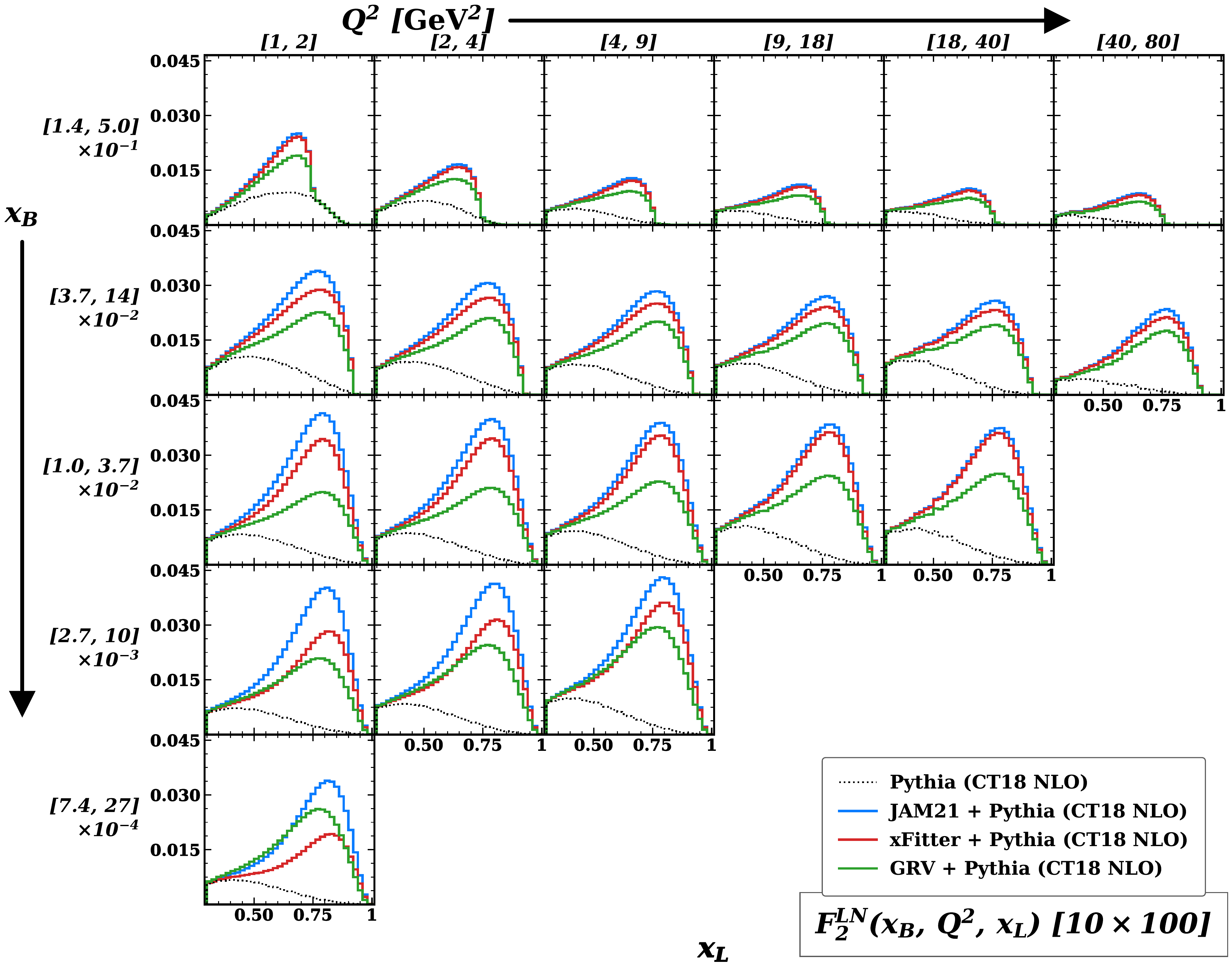}
\caption{Same as Fig.~\ref{fig:eic541}, but for the
  $10\times100~\text{GeV}$ EIC beam energy configuration.}
\label{fig:eic10100}
\end{figure*}

\begin{figure*}[t]
\centering
\includegraphics[width=\textwidth]{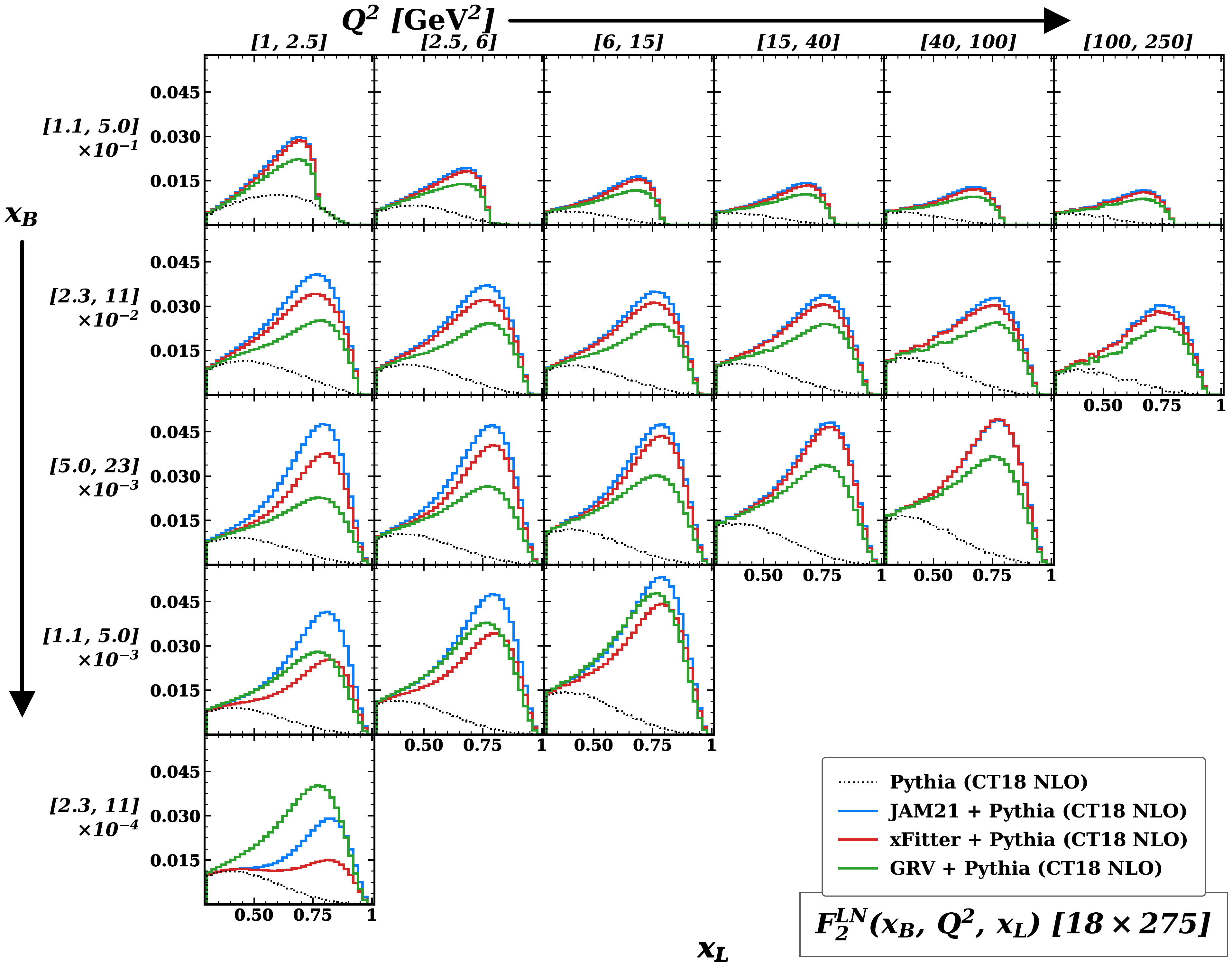}
\caption{Same as Fig.~\ref{fig:eic541}, but for the
  $18\times276~\text{GeV}$ EIC beam energy configuration.}
\label{fig:eic18275}
\end{figure*}

Figures~\ref{fig:eic541}, \ref{fig:eic10100}, and \ref{fig:eic18275}
present the predicted three-fold leading-neutron structure function
$F_2^{\mathrm{LN}(3)}(x_B,Q^2,x_L)$ as a function of $x_L$ across the
multi-differential $(x_B,Q^2)$ kinematic grids for the $5\times41$,
$10\times100$, and $18\times275~\mathrm{GeV}$ EIC configurations,
respectively. The $x_B$ and $Q^2$ bin boundaries, listed in
Table~\ref{tab:eic}, are chosen to span the accessible fiducial phase
space rather than to optimize the statistical precision of individual
bins. The total predictions combine the \textsc{Pythia}~8 DIS
target-fragmentation contribution, evaluated with CT18NLO, with the
OPE Sullivan contribution. The latter is calculated using the JAM21,
xFitter, and GRV pion PDF sets together with the $s$-dependent
exponential pion flux. The forward-neutron angular acceptance is set
to $\theta_n<5$, 2.5, and 1~mrad for the $5\times41$, $10\times100$,
and $18\times275~\mathrm{GeV}$ EIC configurations, respectively, in
order to reduce the DIS target-fragmentation contribution.

QCD evolution suppresses the large-$x_\pi$ contribution and enhances
the small-$x_\pi$ contribution to the Sullivan process as $Q^2$
increases, reflecting the scale dependence of the pion quark
distributions. Consequently, the sensitivity of the Sullivan-process
predictions to the choice of pion PDF becomes more pronounced at small
$x_\pi$, corresponding to smaller $x_B$ at fixed $x_L$. The relative
importance of the DIS target-fragmentation contribution also depends
on the beam-energy configuration and is generally reduced at lower
center-of-mass energies for the same $x_B$ under the acceptance
requirements considered here.

These effects lead to a systematic variation in the relative
contributions of DIS target fragmentation and OPE across the
$(x_B,Q^2)$ kinematic grids. At larger $x_B$ and lower $Q^2$,
corresponding approximately to the upper-left region of each grid, the
two contributions can be of comparable magnitude, such that the total
prediction exceeds the fragmentation component only moderately. In
this region, the sizable target-fragmentation contribution reduces the
relative sensitivity of the leading-neutron yield to pion exchange and
therefore complicates the extraction of pion structure.

Toward smaller $x_B$ and larger $Q^2$, the relative Sullivan-process
contribution increases and can become substantially larger than the
modeled target-fragmentation component. These kinematic regions are
therefore particularly favorable for studies of pion structure, since
the sensitivity to the pion PDFs is enhanced while the relative
contamination from DIS target fragmentation is reduced. The extended
$Q^2$ lever arm also provides sensitivity to the scale dependence of
the pion PDFs and hence to their QCD evolution.

The three beam-energy configurations provide complementary coverage of
this kinematic region. For the $5\times41~\mathrm{GeV}$ setting shown
in Fig.~\ref{fig:eic541}, the accessible range extends to
approximately $Q^2\sim20~\mathrm{GeV}^2$ and $x_B\sim10^{-3}$. The
$10\times100~\mathrm{GeV}$ configuration in Fig.~\ref{fig:eic10100}
extends the coverage to $Q^2\sim80~\mathrm{GeV}^2$ and
$x_B\sim2\times10^{-4}$, providing a broader lever arm for probing the
pion PDFs at intermediate and small momentum fractions. The
$18\times275~\mathrm{GeV}$ configuration shown in
Fig.~\ref{fig:eic18275} provides the broadest kinematic reach,
extending to $Q^2\sim250~\mathrm{GeV}^2$ and
$x_B\sim5\times10^{-5}$. This configuration offers access to
substantially smaller $x_\pi$, where the pion sea-quark distributions
are poorly constrained and the gluon distribution can be probed
indirectly through scaling violations.

As discussed above, an appropriate upper cut on the neutron transverse
momentum can effectively reduce the DIS target-fragmentation
contribution. More restrictive transverse-momentum cuts are required
at higher beam energies to achieve a comparable level of
suppression. With suitably chosen forward-neutron selections, all
three EIC energy configurations can therefore provide useful
sensitivity to the Sullivan process. Taken together, these
measurements offer complementary coverage of the small- and
intermediate-$x_\pi$ regions, bridging the kinematic regimes explored
by HERA leading-neutron measurements and fixed-target pion-induced
Drell--Yan experiments.


\section{Summary and Outlook}
\label{sec:summary}

In this work, we systematically investigated leading-neutron DIS over
the full measured $x_L$ range by combining calculations of the OPE
Sullivan process with DIS target fragmentation modeled using
\textsc{Pythia}~8. Without introducing additional ad hoc empirical
normalization factors, we find that the sum of these two contributions
provides a qualitatively reasonable description of the H1 and ZEUS
data over their measured $(x_B,Q^2,x_L)$ phase space. Some tension
remains in the choice of target-fragmentation model that provides the
best description of the two data sets. Our results suggest that
incorporating data over the full measured $x_L$ range, rather than
restricting global pion PDF analyses to the largest-$x_L$ region,
could provide additional constraints on pion structure. In particular,
the lower-$x_L$ measurements extend the sensitivity of leading-neutron
DIS to smaller $x_\pi$, where the current pion sea-quark and gluon
distributions remain poorly constrained and exhibit substantial PDF
dependence.

Furthermore, our analysis reveals important model dependence in both
the DIS target-fragmentation and pion-flux descriptions. The
forward-physics-tuned beam-remnant configuration is less favored by
the H1 and ZEUS data than several of the other \textsc{Pythia}
settings considered, while the $s$-dependent exponential pion-flux
parameterization substantially underestimates the ZEUS measurements in
the highest-$x_L$ region within the present implementation. The latter
observation appears to differ from the results of previous JAM pion
PDF analyses, which employed the same functional form for the pion
flux.

Our EIC projections demonstrate that the relative importance of the
Sullivan process and DIS target fragmentation varies systematically
across the $(x_B,Q^2)$ phase space and with the beam-energy
configuration. The sensitivity to the pion PDFs increases toward
smaller $x_\pi$ and larger $Q^2$, where QCD evolution enhances the
discrimination among different pion PDF sets, while appropriately
chosen forward-neutron transverse-momentum cuts can suppress the
target-fragmentation contribution. The three EIC beam-energy
configurations therefore provide complementary capabilities: lower
energies offer favorable conditions for controlling fragmentation
backgrounds, whereas higher energies extend the reach to substantially
smaller $x_\pi$ and larger $Q^2$, providing sensitivity to the poorly
constrained pion sea-quark and gluon distributions. Measurements
across multiple EIC energies will thus enable broad and complementary
constraints on pion structure, bridging the kinematic regions explored
by HERA leading-neutron DIS and fixed-target pion-induced Drell--Yan
experiments.

Our work represents an early step toward extending the impact of
leading-neutron DIS measurements on the determination of pion PDFs. A
reliable extraction of pion structure requires a controlled separation
of the OPE Sullivan contribution from competing production mechanisms,
together with a quantitatively reliable description of each
component. Further progress will therefore require improved treatments
of DIS target fragmentation, a careful assessment of the pion--nucleon
form factor and pion-flux parameterization entering the OPE
contribution, and a better understanding of other non-OPE
effects. Complementary constraints from other Sullivan-process
observables, including the nucleon $\bar d/\bar u$
asymmetry~\cite{McKenney:2015xis}, will also be important for testing
the universality of the underlying pion-cloud description. These
developments will help establish a consistent framework in which
leading-neutron DIS data over a broad $x_L$ range can be incorporated
into future global analyses of pion structure.

\section*{Acknowledgment}

We thank Patrick C. Barry for helpful discussions and for providing
information on the JAM pion PDFs. We also thank the Meson Structure
Functions Working Group of the ePIC Collaboration for valuable
comments and suggestions. This work was supported in part by the
National Science and Technology Council of Taiwan (R.O.C.).

\clearpage
\bibliography{ref} 
\newpage
\section*{Appendix}

\subsection{Pion PDF Dependence of Data-to-Theory Structure Function Ratios}
\label{sec:appendix_ratios}

\begin{figure*}[!htbp]
  \centering
  \includegraphics[width=\textwidth]{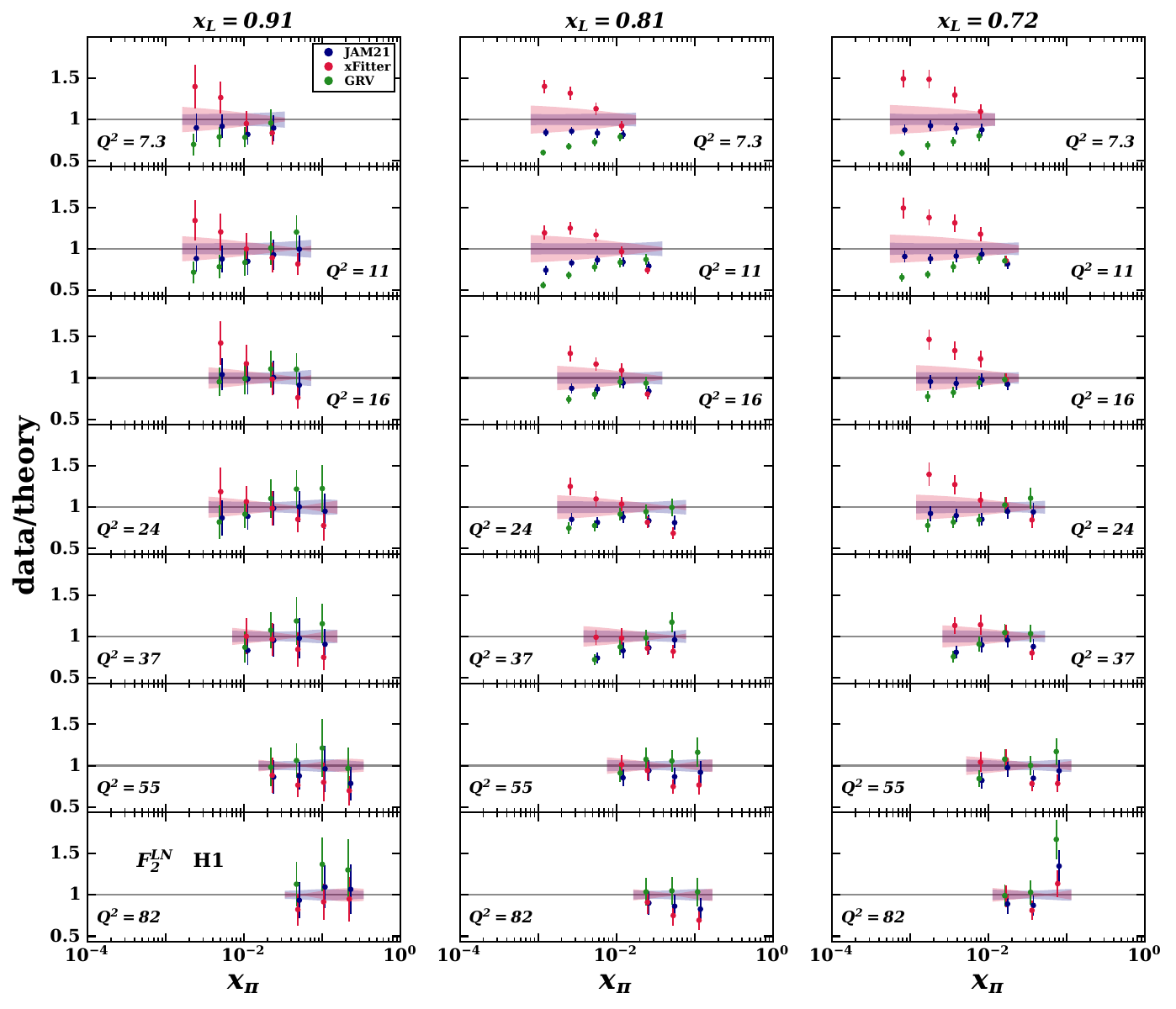}
  \caption{Data-to-theory ratios of the reduced structure function
    $F_2^{\text{LN}(3)}$ as a function of $x_\pi$ for the H1
    experiment~\cite{H1:2010hym} in bins of $Q^2$ and $x_L$ ($x_L =
    0.91, 0.81, 0.72$). Theoretical predictions utilize the JAM21,
    xFitter, and GRV pion PDF sets combined with the $s$-dependent
    exponential pion flux parameterization from
    Eq.~(\ref{eq:regulators}). Shaded bands indicate $1\sigma$ PDF
    uncertainties.}
  \label{fig:h1threedata}
\end{figure*}

\begin{figure*}[!htbp]
  \centering
  \includegraphics[width=\textwidth]{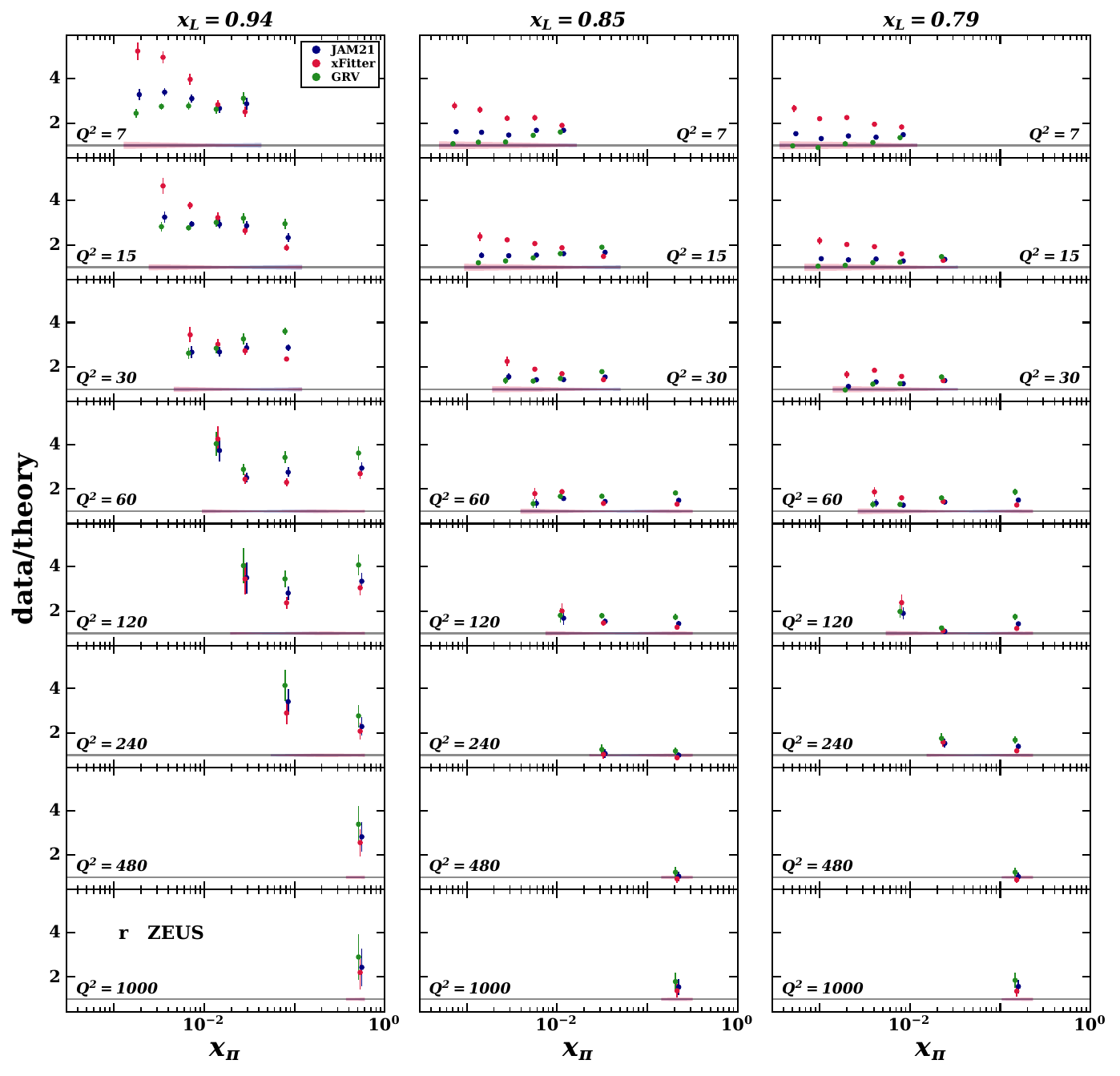}
  \caption{Data-to-theory ratios of the reduced structure function
    $F_2^{\text{LN}(3)}$ as a function of $x_\pi$ for the ZEUS
    experiment~\cite{ZEUS:2002gig} in bins of $Q^2$ and $x_L$ ($x_L =
    0.94, 0.85, 0.79$). Theoretical predictions utilize the JAM21,
    xFitter, and GRV pion PDF sets combined with the $s$-dependent
    exponential pion flux parameterization from
    Eq.~(\ref{eq:regulators}). The CT18NLO proton PDF set is used to
    evaluate $F_2^p(x_B, Q^2)$ for converting the measured $r$ ratios
    into $F_2^{\text{LN}(3)}$. Shaded bands indicate $1\sigma$ PDF
    uncertainties.}
  \label{fig:zeusthreedata}
\end{figure*}

To further quantify the agreement between the large-$x_L$ H1 and ZEUS
measurements and the corresponding Sullivan-process calculations
obtained with different pion PDF sets, Figs.~\ref{fig:h1threedata} and
\ref{fig:zeusthreedata} show the data-to-theory ratios of the
three-fold leading-neutron structure function,
$F_2^{\mathrm{LN}(3),\mathrm{exp}}/F_2^{\mathrm{LN}(3),\mathrm{th}}$,
for the H1 and ZEUS data sets, respectively.

\subsection{Default \textsc{Pythia} settings for DIS target fragmentation}
\label{sec:appendix_pythia}

The default \textsc{Pythia}~8 settings used in this work to study DIS
target fragmentation are listed in Table~\ref{tab:pythia}. They employ
the leading-order NNPDF2.3 proton PDF set~\cite{Ball:2013hta},
together with the default beam-remnant and color-reconnection
settings~\cite{Pythia8317Manual}:
\begin{lstlisting}[basicstyle=\ttfamily\footnotesize, xleftmargin=1pt]
PDF:pSet                          = NNPDF23_lo_as_0130_qed

BeamRemnants:remnantMode          = 0
ColourReconnection:mode           = 0
ColourReconnection:allowJunctions = on

BeamRemnants:dampPopcorn          = 1.0
BeamRemnants:hardRemnantBaryon    = off
BeamRemnants:aRemnantBaryon       = 0.0
BeamRemnants:bRemnantBaryon       = 2.0

BeamRemnants:primordialKTsoft     = 0.9
BeamRemnants:primordialKTremnant  = 0.4
\end{lstlisting}

\subsection{DIS Target Fragmentation Model and Pion PDF Dependence of leading-neutron Structure Function}
\label{sec:appendix_F2Ntotal}

\begin{figure*}[!htbp]
\centering
\includegraphics[width=\textwidth]{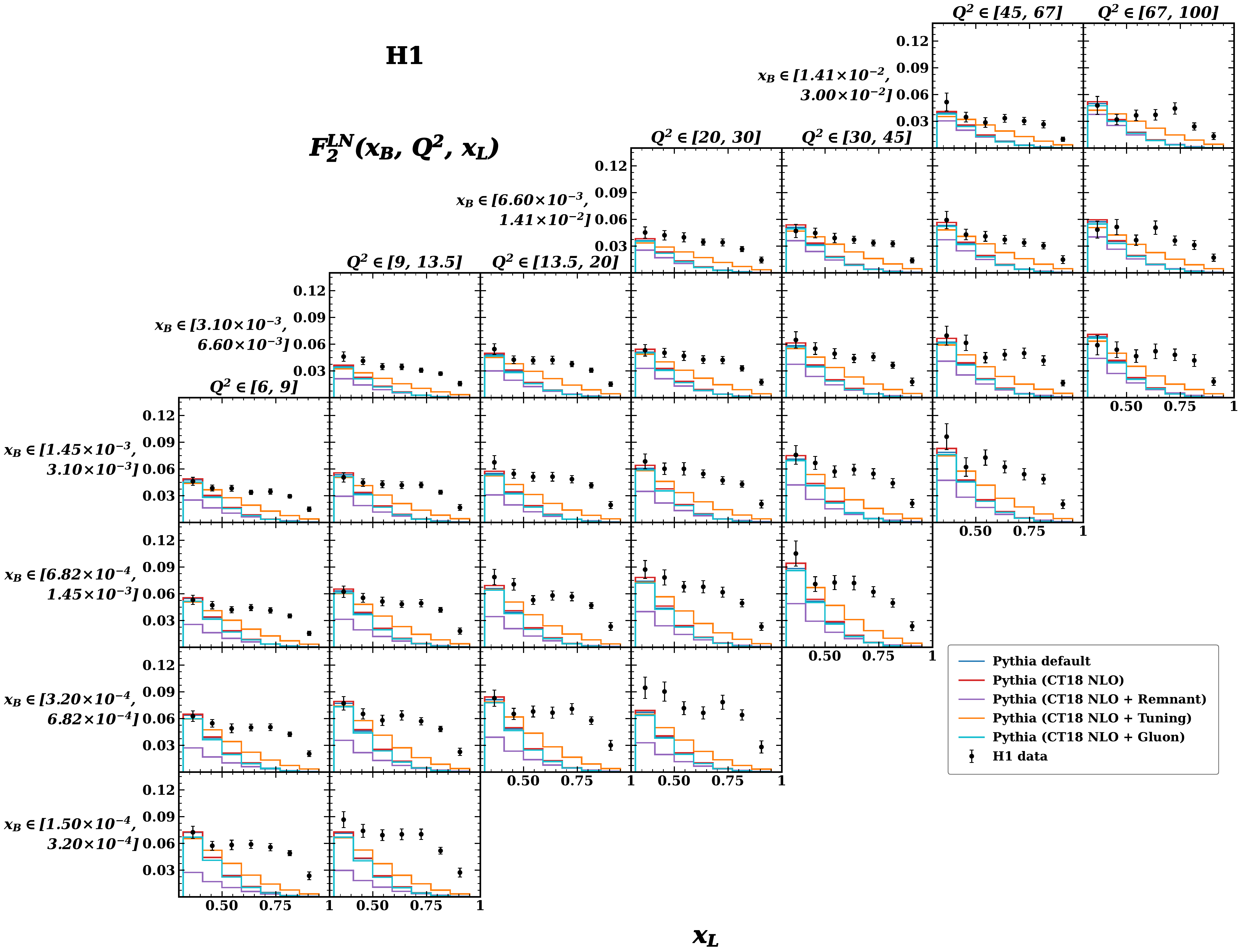}
\caption{The DIS target fragmentation background contribution to the
  leading-neutron structure function
  $F_{2}^{\text{LN}(3)}(x_{B},Q^{2},x_{L})$ plotted versus $x_{L}$
  ($p_{T} \le 0.2~\text{GeV}$) in comparison with H1
  data~\cite{H1:2010hym}. Predictions are shown for various DIS target
  fragmentation schemes implemented in \textsc{Pythia}~8. Columns and
  rows denote fixed $Q^{2}$ and $x_{B}$ bins, respectively.}
\label{fig:f2lnbackground_h1}
\end{figure*}

\begin{figure*}[!htbp]
\centering
\includegraphics[width=\textwidth]{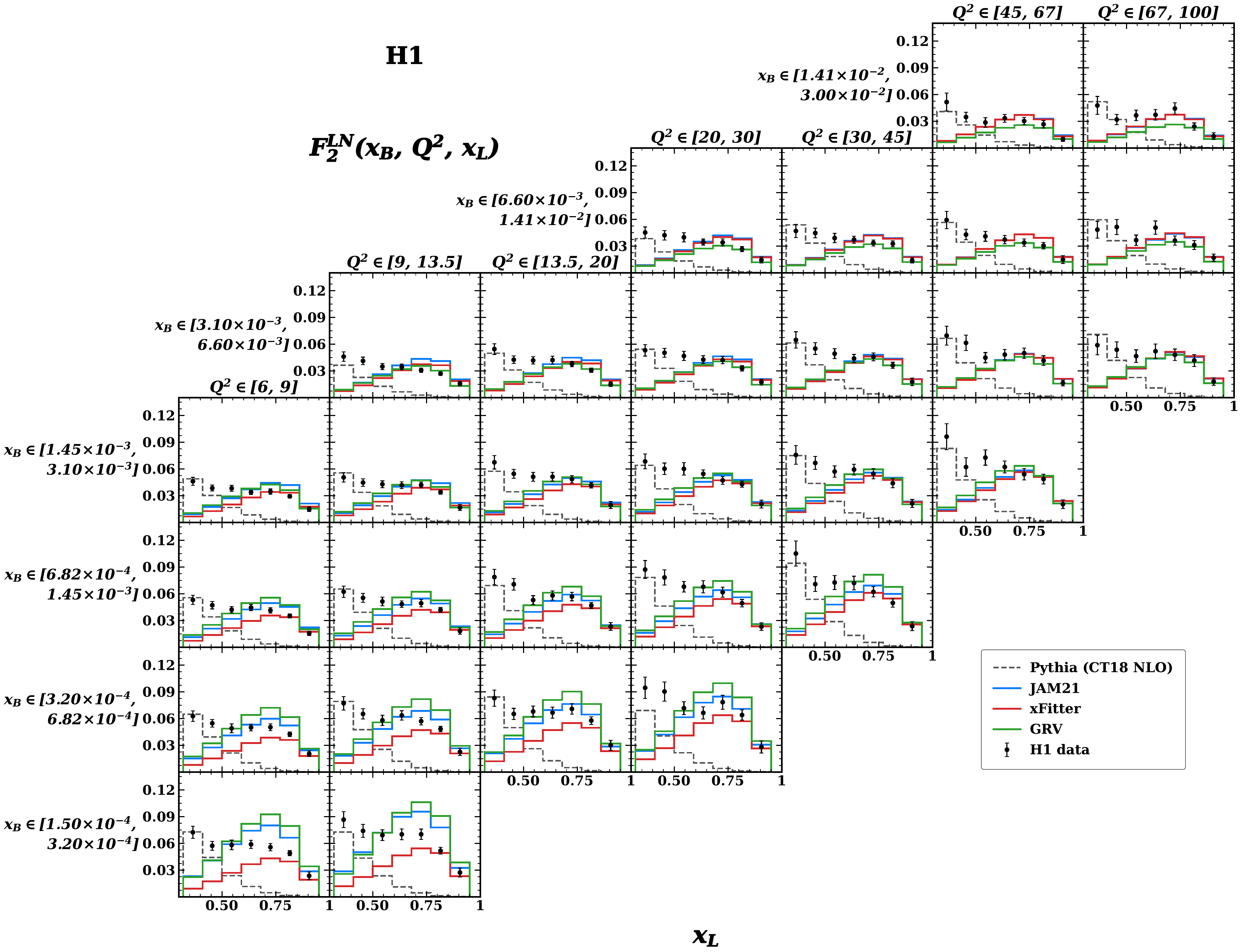}
\caption{Individual DIS target fragmentation and OPE Sullivan-process
  contributions to the leading-neutron structure function
  $F_{2}^{\text{LN}}(x_{B},Q^{2},x_{L})$ in comparison with H1
  data~\cite{H1:2010hym}. Results show the DIS target fragmentation
  contribution generated using default \textsc{Pythia}~8 (with CT18NLO
  proton PDFs), alongside the individual OPE Sullivan contributions
  computed using the JAM21, xFitter, and GRV pion PDF sets.}
\label{fig:f2lnindividual_h1}
\end{figure*}

\begin{figure*}[!htbp]
\centering
\includegraphics[width=\textwidth]{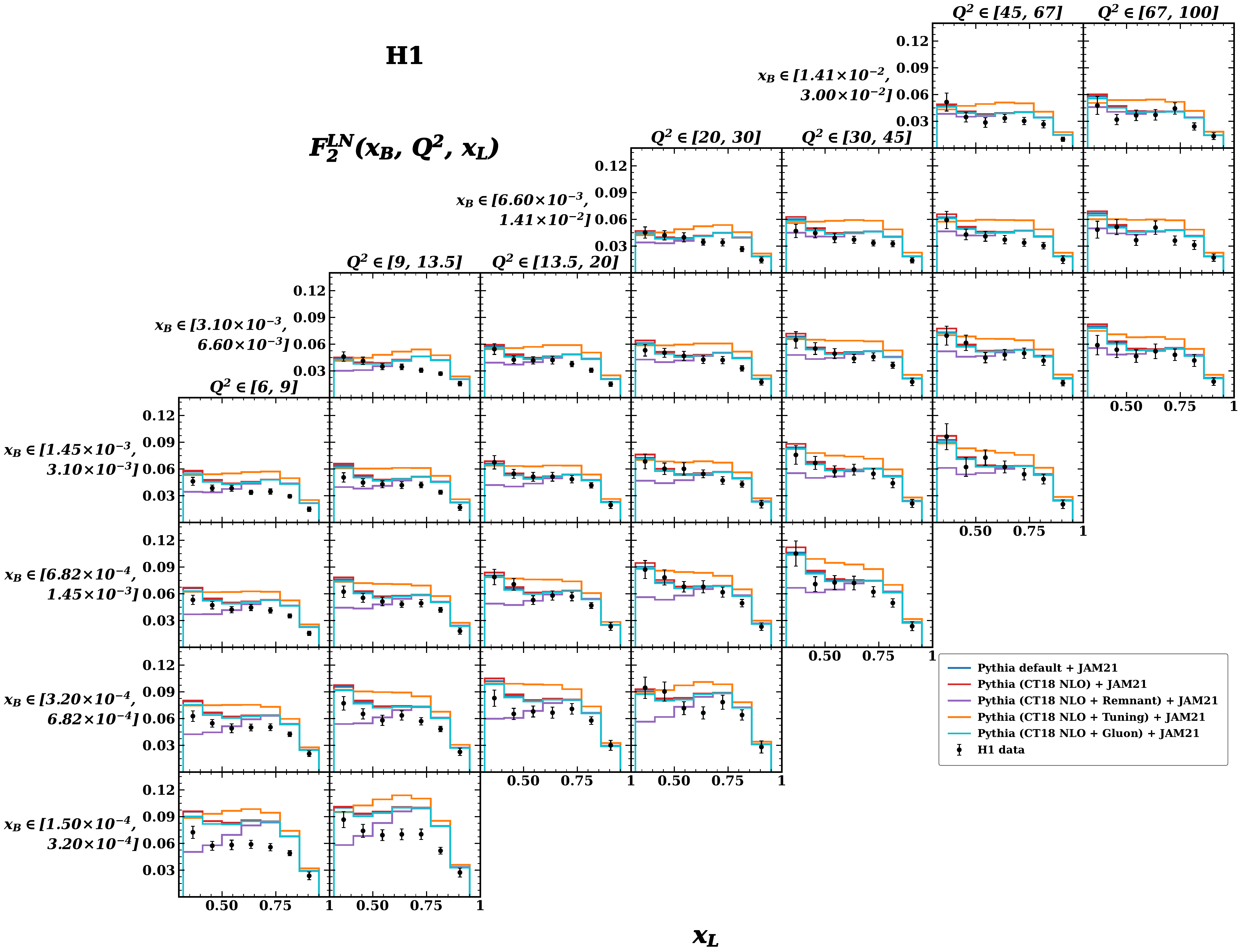}
\caption{The predicted total semi-inclusive leading-neutron structure
  function $F_{2}^{\text{LN}}(x_{B},Q^{2},x_{L})$ as a function of
  $x_{L}$ in comparison with H1 data~\cite{H1:2010hym}. Results are
  obtained by combining the OPE Sullivan contribution, evaluated using
  the JAM21 pion PDF set, with the DIS target fragmentation
  contributions generated using the various \textsc{Pythia}~8 settings
  listed in Table~\ref{tab:pythia}.}
\label{fig:f2lntotal_h1}
\end{figure*}

\begin{figure*}[!htbp]
  \centering
  \includegraphics[width=\textwidth]{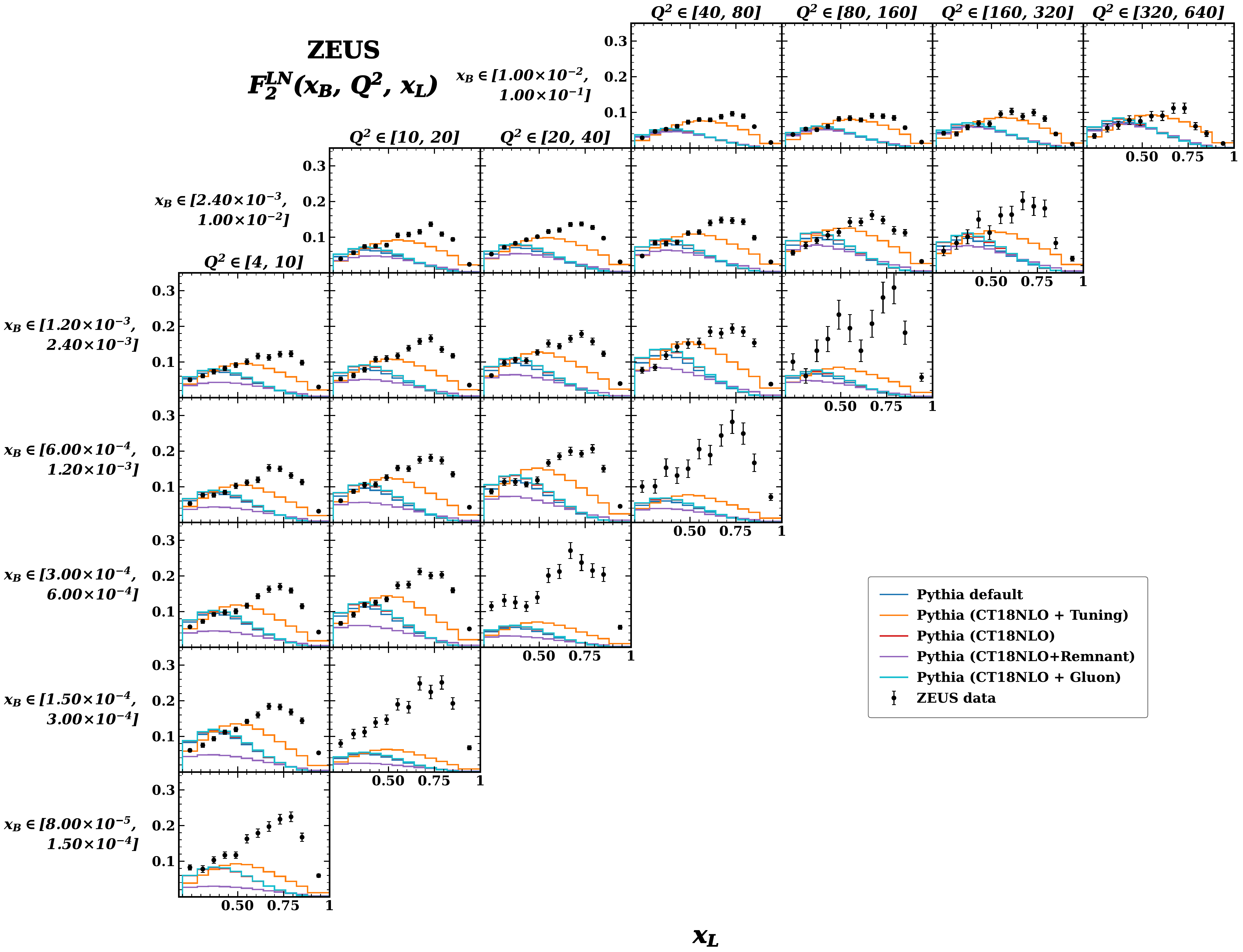}
  \caption{Same as Fig.~\ref{fig:f2lnbackground_h1} for ZEUS
    data~\cite{ZEUS:2002gig}.}
  \label{fig:f2lnbackground_zeus}
\end{figure*}

\begin{figure*}[!htbp]
  \includegraphics[width=\textwidth]{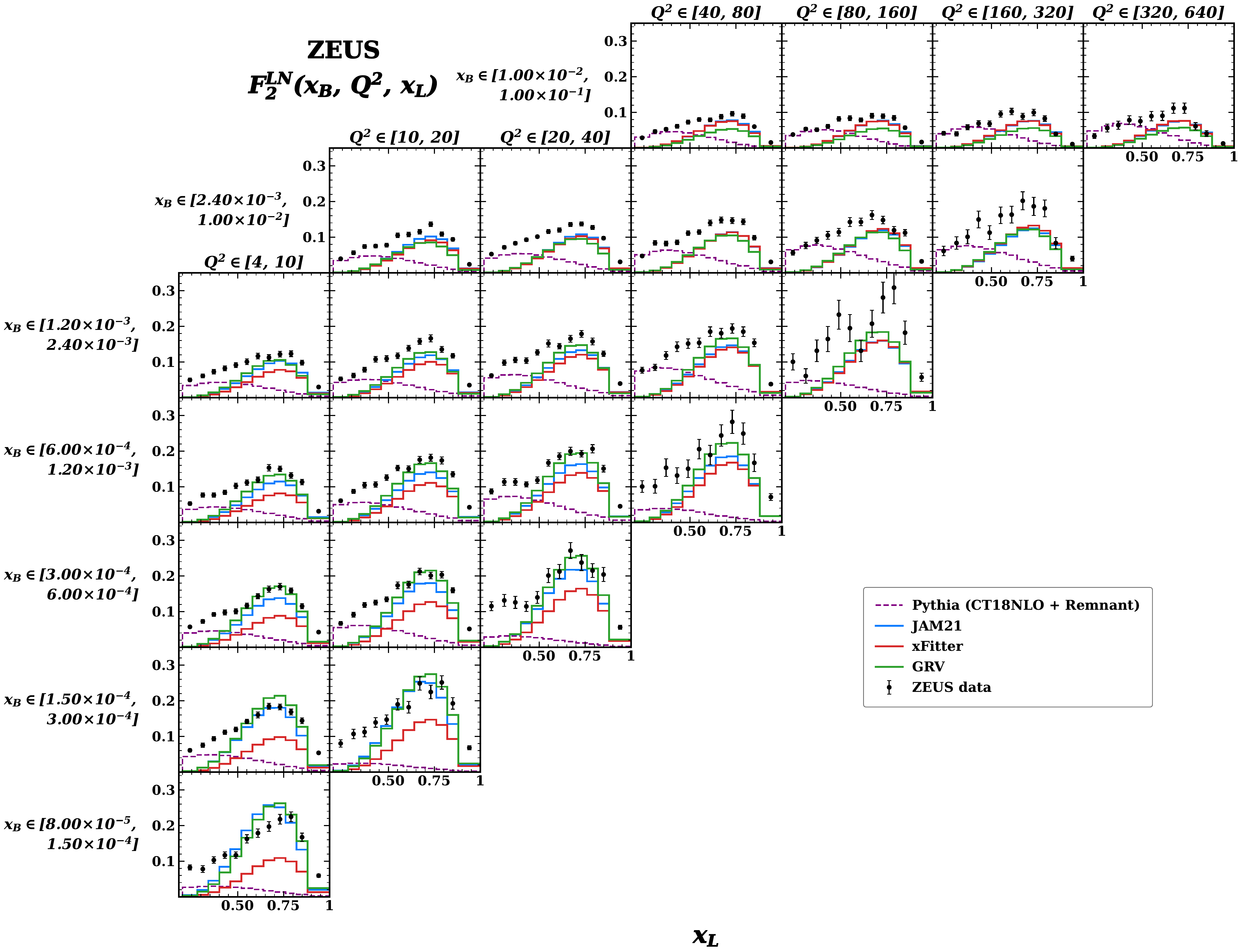}
  \caption{Same as Fig.~\ref{fig:f2lnindividual_h1} for ZEUS
    data~\cite{ZEUS:2002gig}.}
  \label{fig:f2lnindividual_zeus}
\end{figure*}

\begin{figure*}[!htbp]
  \includegraphics[width=\textwidth]{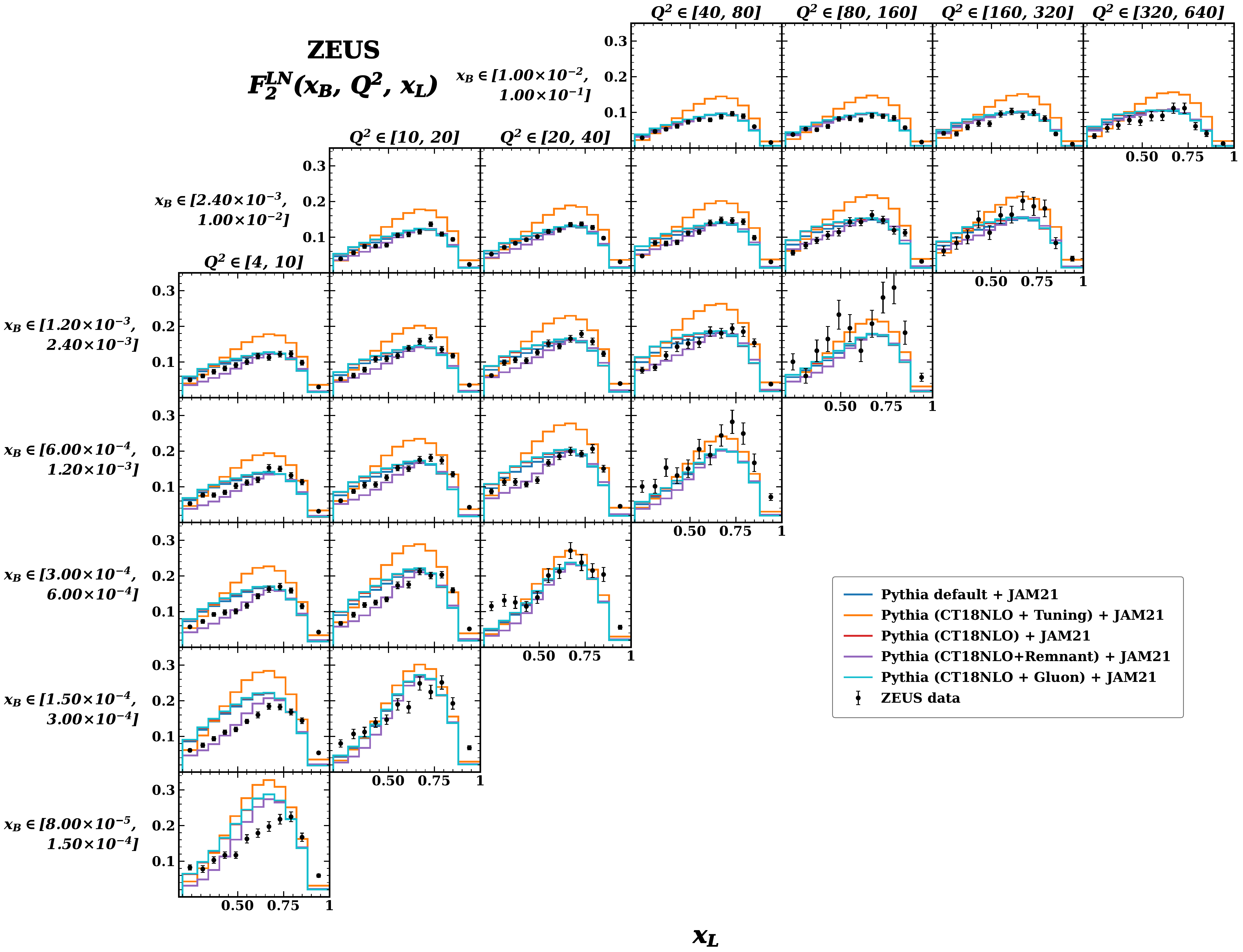}
  \caption{Same as Fig.~\ref{fig:f2lntotal_h1} for ZEUS
    data~\cite{ZEUS:2002gig}.}
  \label{fig:f2lntotal_zeus}
\end{figure*}

To systematically investigate the interplay between DIS target
fragmentation and the one-pion-exchange (OPE) contribution,
Figs.~\ref{fig:f2lnbackground_h1}--\ref{fig:f2lntotal_zeus} compare
the predicted three-fold leading-neutron structure function
$F_2^{\mathrm{LN}(3)}(x_B,Q^2,x_L)$ with the H1 and ZEUS measurements.

For the H1 data, Fig.~\ref{fig:f2lnbackground_h1} shows the DIS
target-fragmentation contribution obtained with the different
\textsc{Pythia}~8 configurations listed in
Table~\ref{tab:pythia}. Figure~\ref{fig:f2lnindividual_h1} compares
the target-fragmentation contribution with the individual OPE
predictions calculated using the JAM21, xFitter, and GRV pion
PDFs. The combined predictions, obtained by adding the JAM21 OPE
contribution to each of the \textsc{Pythia}~8 fragmentation
configurations, are compared with the H1 measurements in
Fig.~\ref{fig:f2lntotal_h1}.

The corresponding comparisons for the ZEUS data are shown in
Figs.~\ref{fig:f2lnbackground_zeus}, \ref{fig:f2lnindividual_zeus},
and \ref{fig:f2lntotal_zeus}, following the same sequence of
target-fragmentation, individual OPE, combined, and global model
comparisons, respectively.

\subsection{Kinematic binning for leading-neutron structure functions at the EIC}
\label{sec:appendix_EIC_kinematic}

Table~\ref{tab:eic} summarizes the kinematic binning adopted for the
differential leading-neutron structure functions for the three EIC
beam-energy configurations. For each configuration, six $Q^2$ bins and
five $x_B$ bins are defined. Bins that violate the physical constraint
$y=Q^2/(s,x_B)\leq1$ or contain insufficient \textsc{Pythia}
statistics are excluded from the analysis.

\begin{table}[!hbtp]
\centering
\caption{Kinematic binning used for the leading-neutron structure-function
grids across the three EIC beam-energy configurations.}
\label{tab:eic}
\renewcommand{\arraystretch}{1.25}
\setlength{\tabcolsep}{4pt}
\footnotesize
\begin{tabular}{l c c}
\hline\hline
\textbf{Configuration} & \textbf{$Q^2$ binning ($\text{GeV}^2$)} & \textbf{$x_B$ binning} \\
\hline
\textbf{$5 \times 41~\text{GeV}$} & Total: $[1, 20]$ & Total: $[2.8\times 10^{-3}, 0.5]$ \\
($6\,Q^2 \times 5\,x_B$ bins) & $[1, 1.5]$ & $[1.8, 5.0] \times 10^{-1}$ \\
 & $[1.5, 2.5]$ & $[6.3, 18] \times 10^{-2}$ \\
 & $[2.5, 4]$ & $[2.2, 6.3] \times 10^{-2}$ \\
 & $[4, 7]$ & $[7.9, 22] \times 10^{-3}$ \\
 & $[7, 12]$ & $[2.8, 7.9] \times 10^{-3}$ \\
 & $[12, 20]$ & \\
\hline
\textbf{$10 \times 100~\text{GeV}$} & Total: $[1, 80]$ & Total: $[7.4\times 10^{-4}, 0.5]$ \\
($6\,Q^2 \times 5\,x_B$ bins) & $[1, 2]$ & $[1.4, 5.0] \times 10^{-1}$ \\
 & $[2, 4]$ & $[3.7, 14] \times 10^{-2}$ \\
 & $[4, 9]$ & $[1.0, 3.7] \times 10^{-2}$ \\
 & $[9, 18]$ & $[2.7, 10] \times 10^{-3}$ \\
 & $[18, 40]$ & $[7.4, 27] \times 10^{-4}$ \\
 & $[40, 80]$ & \\
\hline
\textbf{$18 \times 275~\text{GeV}$} & Total: $[1, 250]$ & Total: $[2.3\times 10^{-4}, 0.5]$ \\
($6\,Q^2 \times 5\,x_B$ bins) & $[1, 2.5]$ & $[1.1, 5.0] \times 10^{-1}$ \\
 & $[2.5, 6]$ & $[2.3, 11] \times 10^{-2}$ \\
 & $[6, 15]$ & $[5.0, 23] \times 10^{-3}$ \\
 & $[15, 40]$ & $[1.1, 5.0] \times 10^{-3}$ \\
 & $[40, 100]$ & $[2.3, 11] \times 10^{-4}$ \\
 & $[100, 250]$ & \\
\hline\hline
\end{tabular}
\end{table}


\end{document}